\documentclass[10pt]{article}
\usepackage{fullpage}
\usepackage{setspace}
\usepackage{parskip}
\usepackage{titlesec}
\usepackage{placeins}
\usepackage{xcolor}
\usepackage{lineno}
\PassOptionsToPackage{hyphens}{url}
\usepackage[colorlinks = true,
            linkcolor = blue,
            urlcolor  = blue,
            citecolor = blue,
            anchorcolor = blue]{hyperref}
\usepackage{etoolbox}
\makeatletter
\usepackage{scicite}
\renewenvironment{abstract}
  {{\bfseries\noindent{\abstractname}\par\nobreak}\footnotesize}
  {\bigskip}

\titlespacing{\section}{0pt}{*3}{*1}
\titlespacing{\subsection}{0pt}{*2}{*0.5}
\titlespacing{\subsubsection}{0pt}{*1.5}{0pt}
\usepackage{authblk}
\usepackage{graphicx}
\usepackage[space]{grffile}
\usepackage{latexsym}
\usepackage{textcomp}
\usepackage{longtable}
\usepackage{tabulary}
\usepackage{booktabs,array,multirow}
\usepackage{amsfonts,amsmath,amssymb}
\providecommand\citet{\cite}
\providecommand\citep{\cite}

\newif\iflatexml\latexmlfalse
\AtBeginDocument{\DeclareGraphicsExtensions{.pdf,.PDF,.eps,.EPS,.png,.PNG,.tif,.TIF,.jpg,.JPG,.jpeg,.JPEG}}
\usepackage[utf8]{inputenc}
\usepackage[english]{babel}
\begin{document}
\title{\textbf{Unveiling the Physics of the  Mutual Interactions in Paramagnets}}
\author[1]{Lucas Squillante}%
\affil[1]{Departamento de F\'isica, S\~ao Paulo State University (UNESP), IGCE, Rio Claro, 13506-900 SP, Brazil }%
\author[1]{Isys F. Mello}
\author[2]{Gabriel O. Gomes}
\affil[2]{Department of Astronomy, University of São Paulo, 05508-090 SP, Brazil}%
\author[3]{A. C. Seridonio}
\affil[3]{S\~ao Paulo State University (Unesp), Department of Physics and Chemistry, Ilha Solteira - SP, Brazil}%
\author[1]{R. E. Lagos-Monaco}
\author[4]{H. Eugene Stanley}
\affil[4]{Boston University, Department of Physics, Boston, 02215, USA}%
\author[1$\ast$]{Mariano de Souza}
\affil[$^\ast$]{\normalsize{To whom correspondence should be addressed. E-mail: mariano.souza@unesp.br}}
\vspace{-1em}
\date{}
\begingroup
\let\center\flushleft
\let\endcenter\endflushleft
\maketitle
\endgroup

\begin{abstract}
{In real paramagnets, there is always a subtle many-body contribution to the system's energy, which can be regarded as a small effective local magnetic field ($B_{loc}$). Usually, it is neglected, since it is very small when compared with thermal fluctuations and/or external magnetic fields ($B$). Nevertheless, as both the temperature $(T) \rightarrow$ 0\,K and $B \rightarrow$ 0\,T, such many-body contributions become ubiquitous. Here, employing the magnetic Gr\"uneisen parameter ($\Gamma_{mag}$) and entropy arguments,  we report on the pivotal role played by the mutual interactions in the regime of ultra-low-$T$ and vanishing $B$. Our key results are:  \emph{i}) absence of a genuine zero-field quantum phase transition due to the presence of $B_{loc}$; \emph{ii}) connection between the canonical definition of temperature and $\Gamma_{mag}$; and \emph{iii}) possibility of performing adiabatic magnetization by only manipulating the mutual interactions. Our findings unveil unprecedented aspects emerging from the mutual interactions.
}\\%
\end{abstract}%

\section*{Introduction}
\vspace{0.3cm}

Magnetic excitations in solids have been broadly investigated in the past decades, being crucial to the understanding of exotic physical phenomena such as superconductivity \cite{Mathur} and magnetic field induced quantum phase transitions \cite{Geibel}, just to mention a few examples. It is well-known that the behavior of paramagnetic metals and insulators are nicely described, respectively, by the Pauli paramagnetism and Brillouin-like model. Such approaches are based on a spin gas scheme, i.e., interactions between magnetic moments are not taken into account and thus the system is treated as an \emph{ideal} paramagnet. However, in \emph{real} paramagnets  the magnetic dipolar interactions between adjacent spins are always present, being usually neglected.  Although the mutual interactions, also called zero-field splitting \cite{kittel,ashcroft} and \emph{champ mol\'eculaire} \cite{weiss, linuspauling}, have been broadly mentioned in the literature  \cite{cryophysics, absolutezero,  madelung, vanvleck, ralph, blundell, tari, pathria, feynmann, pobel, fazekas}, a detailed discussion about their role in the magnetic properties of solids in the regime of ultra low-temperatures and vanishing external magnetic field is still lacking.
\begin{figure}[h!]
\centering
\includegraphics[width=\textwidth]{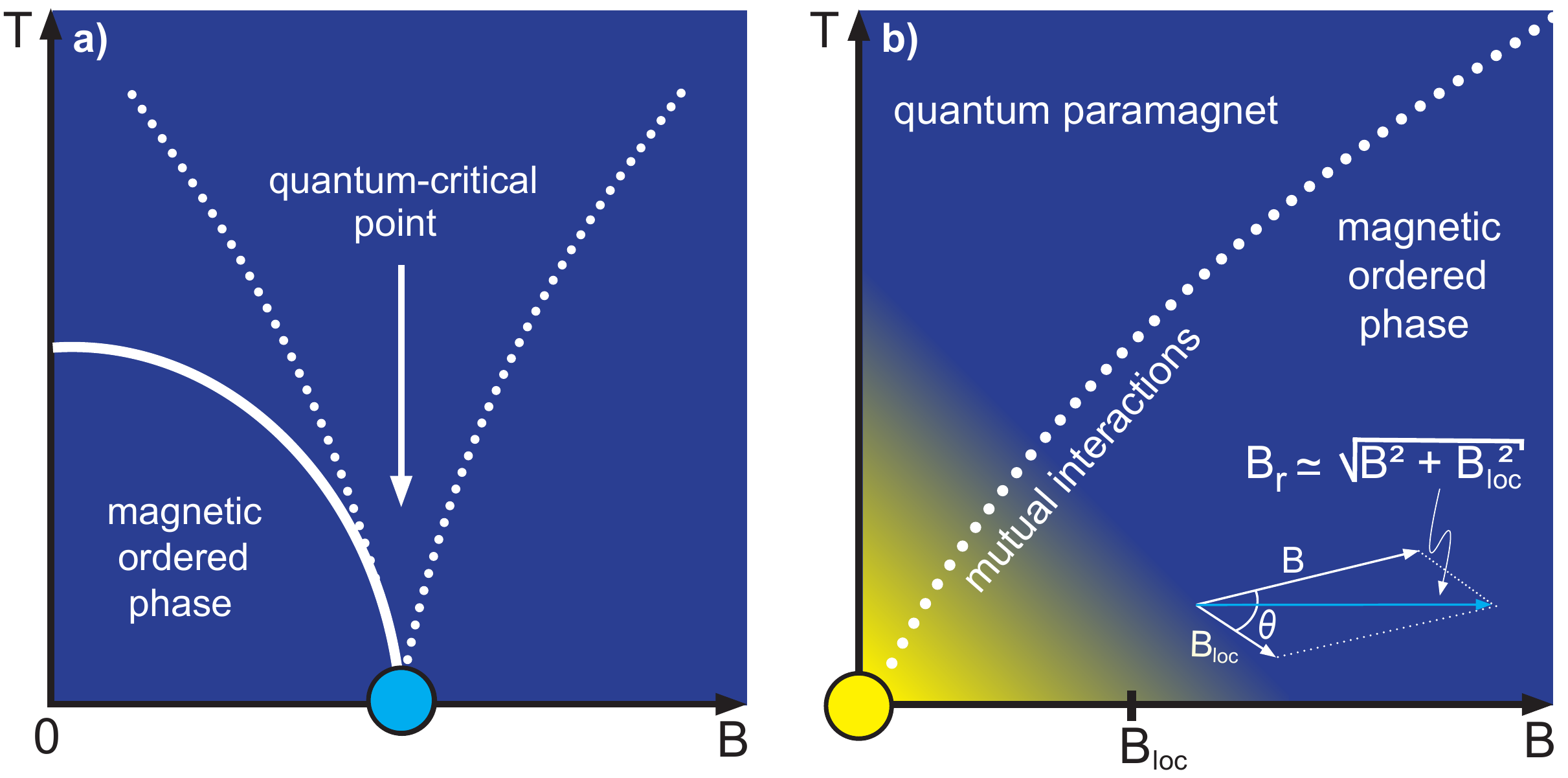}
\caption{\footnotesize \textbf{External magnetic field-induced quantum phase transitions}. Schematic temperature $T$ \emph{versus} external magnetic field $B$ phase diagrams. a) Magnetic field-induced (cyan bullet) quantum critical point (QCP). The dotted lines depict the crossover temperatures. Figure after Ref.\cite{zhu}. b) Hypothetical zero-field quantum critical point (yellow bullet). The yellow gradient shadow represents the role played by the mutual interactions between neighboring magnetic moments, which are responsible for the emergence of an effective local magnetic field $B_{loc}$. We consider that $B_{loc}$ makes an angle $\theta \simeq$ 90$^{\circ}$ with $B$,  so that the resultant magnetic field (inset) is given by $B_r \simeq \sqrt{B^2 + {B_{loc}}^2}$ (cyan arrow). The dotted line represents the transition from quantum paramagnet to a magnetic ordered phase. Details in the main text.}
\label{Fig-1}
\end{figure}
Nowadays, exotic manifestations of matter, like non-Fermi-liquid behavior and unconventional superconductivity \cite{Putzke, Huang, kanoda, cuprates}, emerging in the immediate vicinity of a quantum critical point (QCP), have been attracting high interest of the community. It is well-established that for a pressure-induced QCP \cite{zhu,gegenwart}, as well as for a finite temperature ($T$) pressure-induced critical point \cite{barto,epj,stanley}, the Gr\"uneisen ratio, i.e., the ratio between thermal expansivity and specific heat at constant pressure, is enhanced upon approaching the critical values of the tuning parameter and it diverges right at the critical point. For a magnetic field-induced QCP \cite{PNAS} the analogous physical quantity to the Gr\"uneisen ratio is the so-called magnetic Gr\"uneisen parameter, hereafter  $\Gamma_{mag}$  \cite{zhu,mgarst,prbmce}. The enhancement of both the Gr\"uneisen ratio and $\Gamma_{mag}$ in the immediate vicinity of a magnetic field-induced QCP is merely a direct consequence of the high entropy accumulation in that region  \cite{zhu,gegenwart}, which in turn is related to the fluctuations of the order parameter. Also, it is well-known that $\Gamma_{mag}$  quantifies the magneto-caloric effect \cite{anders,Moya}, which in turn enables to change the temperature of a system upon varying adiabatically the external applied magnetic field \cite{mgarst,prbmce,zhu}. The fingerprints of a genuine magnetic field-induced QCP, besides the gradual suppression of an order parameter (or energy scale) at the QCP [Fig.\,\ref{Fig-1} a)], are: \emph{i}) the divergence of $\Gamma_{mag}$ for $T \rightarrow $ 0\,K at the critical magnetic field $B_c$; \emph{ii}) the sign-change of $\Gamma_{mag}$ upon crossing $B_c$, and \emph{iii}) its typical scaling behavior in the form $T/(B - B_c)^{\epsilon}$, where $B$ is the external applied magnetic field and $\epsilon$ the scaling exponent \cite{heavyfermion, gegenwart}.
Particular attention has been paid to the so-called zero-field quantum criticality [Fig.\,\ref{Fig-1} b)], i.e., the system is inherently quantum critical ($B_c$ = 0\,T) and thus no field-sweep is required for achieving the QCP.  Examples include several materials, such as YbCo$_2$Ge$_4$ \cite{sakai}, $\beta$-YbAlB$_4$ \cite{yosuke}, and Au-Al-Yb \cite{deguchi}. In the case of $\beta$-YbAlB$_4$, in particular, the understanding of possible zero-field quantum criticality \cite{yosuke,ramires2012,tomita2015} and the emergence of superconductivity at $T_c =$ 80\,mK \cite{kuga2008,nakatsuji2008} remains elusive \cite{coleman2016,tomita2015b}. Interestingly, although $\beta$-YbAlB$_4$ is metallic \cite{kuga2008,nakatsuji2008},  its magnetic susceptibility displays a typical Curie-Weiss-like behavior \cite{yosuke} and it is thus evident that resultant local magnetic moments are still present into the system at very low temperatures, cf.\,Fig.\,2(A) of Ref.\,\cite{yosuke}. Because of such exotic behavior the system is considered as a ``\emph{strange metal}'' \cite{gegenwart}. Based on a scaling analysis of the magnetization,  the authors of Ref.\,\cite{yosuke} argue on the crossover of $\beta$-YbAlB$_4$ from a non Fermi liquid to a Fermi liquid behavior.
Recently, some of us reported on a surprisingly divergent behavior of $\Gamma_{mag}$  for model systems, including the one-dimensional Ising model under longitudinal $B$ and Brillouin-like paramagnets \cite{prbmce}.
Here, we report on the  absence of zero-field quantum criticality for \emph{any} paramagnetic insulator with non-zero effective local magnetic field ($B_{loc}$) and discuss the intricate role played by mutual interactions in the regime of $T \rightarrow$ 0\,K and $B \rightarrow$\,0\,T.  We demonstrate the validity of our analysis for the textbook Brillouin-like paramagnet and for the proposed zero-field quantum critical system  $\beta$-YbAlB$_4$.
At some extent, our approach is reminiscent of the famous Mermin-Wagner theorem \cite{Mermin}, since at finite temperatures the mutual interactions give rise to long-range intrinsic magnetic fluctuations.
Also, the connection between $\Gamma_{mag}$ and the canonical definition of temperature is reported.  Yet, we propose the possibility of carrying out adiabatic magnetization by only manipulating the mutual interactions. It is to be noted that a discussion in the literature concerning on adiabatic magnetization was reported about seventy years ago  \cite{Wolf}, when adiabatic magnetization was employed to produce cooling using paramagnetic salts with $(\partial S/\partial B)_T > 0$, where $S$ is the entropy.
When dealing with possible zero-field quantum criticality in paramagnets, we need to be very careful by analyzing the divergence of $\Gamma_{mag}$ for $B \rightarrow$ 0\,T experimentally, since an enhancement of $\Gamma_{mag}$ solely does not suffice to assign genuine zero-field quantum criticality \cite{prbmce}.
It turns out that due to the mutual interactions a spontaneous magnetically ordered phase emerges, which prevents $\Gamma_{mag}$ to diverge.
For the sake of completeness, it is worth recalling that paramagnetic systems have been used to achieve low-temperatures in the range of $\mu$K (electronic spins) and nK (nuclear spins) employing the adiabatic demagnetization method \cite{ralph}. Nevertheless, the achievement of exactly 0\,K using this process is limited by the increase of the mutual interactions' relevance upon approaching the ground-state [Fig.\,\ref{Fig-1} b)].
In quantitative terms, considering that in an ideal paramagnet neighboring magnetic moments $\vec {\mu}$ are separated by a distance $r$ \cite{odom}, the magnetic dipolar energy is roughly given by $U_{dip} \simeq (\mu_0\mu^2)/(4\pi r^3) $\cite{blundell}, where $\mu_0 =$ (4$\pi$$\times$10$^{-7}$)\,T$\cdot$m/A is the vacuum magnetic permeability. Since classically the magnetic energy is given by $U_{mag}$ = $-\mu B_{loc} cos\phi$, being $\phi$ the angle between $\vec{\mu}$ and $\vec{B}_{loc}$, the local dipolar magnetic field is roughly given by $B_{loc} \simeq \mu_0\mu/4\pi r^3$. Hence, for $r$ = 5$\textmd{\AA}$, a typical distance between neighboring spins in a paramagnet, and considering $\mu = \mu_B =$ (9.27$\times$10$^{-24}$)\,J/T, an intrinsic effective local magnetic field $B_{loc} \simeq$ 0.01\,T can be estimated \cite{ralph}.
It is worth emphasizing that although we have considered that $B_{loc}$ emerges purely from the magnetic dipolar interactions between neighboring magnetic moments, it is clear that the electrostatic energy is also present into the system. Nevertheless, the electrostatic energy overcomes the magnetic energy only in the regime of relatively high temperatures \cite{Reif}. Considering that we are interested in the physical properties of paramagnets in the ultra-low $T$ ($<<$ 1\,K) regime our analysis of $B_{loc}$ for real systems remains appropriate, when only the magnetic dipolar interactions are considered.
If we consider the Hydrogen atom, for instance, we must take into account the interaction between electronic and nuclear spins, i.e.\,, the hyperfine coupling, so that when the distance between electron and nucleus is the Bohr radius, the electron perceives a local magnetic field from the nuclear spin $B_{loc} \simeq$ 0.0063\,T. The latter is roughly one order of magnitude lower than typical values of  $B_{loc}$ in real paramagnets as expected, since the nuclear magnetic moment is roughly 1000 times lower than the electronic one.
In the frame of Quantum Mechanics, the Eigenenergies of the Hydrogen for the Zeeman splitting are obtained through the following Hamiltonian \cite{feynmann}:

\begin{equation}
\widehat{H} = A(\boldsymbol{\sigma^e} \cdot \boldsymbol{\sigma^p})-\mu_e \boldsymbol{\sigma^e} \cdot \textbf{B}-\mu_p \boldsymbol{\sigma^p} \cdot \textbf{B},
\label{feynman}
\end{equation}

where the indexes $e$ and $p$ refers to electron and proton, respectively; $\sigma$ is the spin operator and $A$ is a constant related to the magnetic interaction between electron and proton \cite{feynmann}. The first term of this simple Hamiltonian (Eq.\,\ref{feynman}) is independent of the external magnetic field and, therefore, it is connected to the zero-field splitting. Indeed, the Zeeman splitting starts at $U_{mag}$ = 0\,J and the magnetic energy difference between the energy levels enhances as the external magnetic field is increased. It turns out that the Eigenenergies of the Hamiltonian (Eq.\,\ref{feynman}) show a quite similar behavior, being that in this case for $B$ = 0\,T the magnetic energy is $U_{mag} = A$. Analogously, if we consider two neighboring magnetic moments, we can treat the mutual interactions between them employing the Hamiltonian (Eq.\,\ref{feynman}), being necessary only to associate them to the indexes $e$ and $p$.  As we have mentioned before, given its relatively low strength, $B_{loc}$ is relevant only for vanishing external applied magnetic fields and in the regime of ultra low-temperatures. In other words,  $B_{loc}$ begins to be important only when the thermal and magnetic energies are comparable, recalling that $U_{mag} = -\vec{\mu} \cdot \vec{B}_{loc}$ and $U_{mag} = + \vec{\mu} \cdot \vec{B}_{loc}$ represent, respectively, the minimum and maximum magnetic energies.
The magnetic energy can also be expressed in terms of the total angular momentum quantum number $J$,  as well as in terms of the magnetic quantum number $m_J$, which represents the number of allowed orientations of $\vec{\mu}$. The modulus of the total angular momentum vector $\vec {J}$ is given in terms of $J$, namely $ |\vec{J}| = \hbar\sqrt{J(J+1)}$, where $\hbar$ is Planck's constant divided by 2$\pi$. Each value of $m_J = -J$, $-J+1$, ..., $J-1$, $J$ describes a particular orientation of the magnetic moment and its respective magnetic energy, since $\mu = g\mu_B m_J$, where $g$ is  the gyromagnetic factor \cite{alberto}.  Here it is worth emphasizing that in our analysis we consider $\vec{\mu}$ oriented along the $\vec{z}$ direction.
In the frame of Quantum Mechanics, the magnetic energy is written as $U_{mag} = -\left[m_J/\sqrt{J(J+1)}\,\right]\mu B_{loc}$ \cite{alberto}.
At this point, we stress that in our analysis $B_{loc}$ is the effective magnetic field generated by the magnetic dipolar interaction between  neighboring magnetic moments. Since $\sqrt{J(J+1)}$ is always greater than $|m_J|$, the magnetic energy is not exactly $\pm$ $\mu B_{loc}$ as expected classically, and thus we can infer that $\phi$ will never be exactly zero or  $\pi$ \cite{alberto}.
In fact, the absence of a perfect alignment between $\vec{\mu}$ and $\vec{B}_{loc}$ can be observed in the results shown in Fig.\,\ref{Fig-2}  for the Brillouin paramagnet (upper panel) and $\beta$-YbAlB$_4$ (lower panel). Note that the entropy is lowered upon increasing $B$, but it will never be exactly zero for any finite value of $B$.
Yet, in quantitative terms, in the case of $\beta$-YbAlB$_4$ \cite{yosuke}, using the effective Yb magnetic moment  $\mu \simeq$ 1.94$\mu_B$ and the Yb-Yb separation of $r \simeq$ 3.5\,$\textmd{\AA}$ \cite{macaluso}, results in $B_{loc}$ $\simeq$ 0.04\,T.
Even considering the fact that the system undergoes a superconducting transition at $T_c$ = 80\,mK \cite{kuga2008,nakatsuji2008}, intrinsic magnetic moments survive in $\beta$-YbAlB$_4$ at very low-$T$ \cite{yosuke}, so that the Yb valence fluctuations \cite{matsumoto2012} do not affect our analysis. We consider that such resultant magnetic moment is responsible for the emergence of $B_{loc}$ and thus, even into the superconducting dome, $B_{loc}$ is relevant and prevents that zero-field quantum criticality takes place.
Interestingly enough, in the case of  $\alpha$-YbAlB$_4$ \cite{kuga2018} the magnetic moments are fully screened and thus it is not possible to infer resultant magnetic dipolar interactions. After this Introduction, we present and discuss our findings.

\section*{Results and Discussion}

\vspace{0.3cm}
\textbf{\emph{i}) Absence of a genuine zero-field quantum phase transition due to the presence of $B_{loc}$.}

\begin{figure}[h!]
\centering
\includegraphics[scale=0.86]{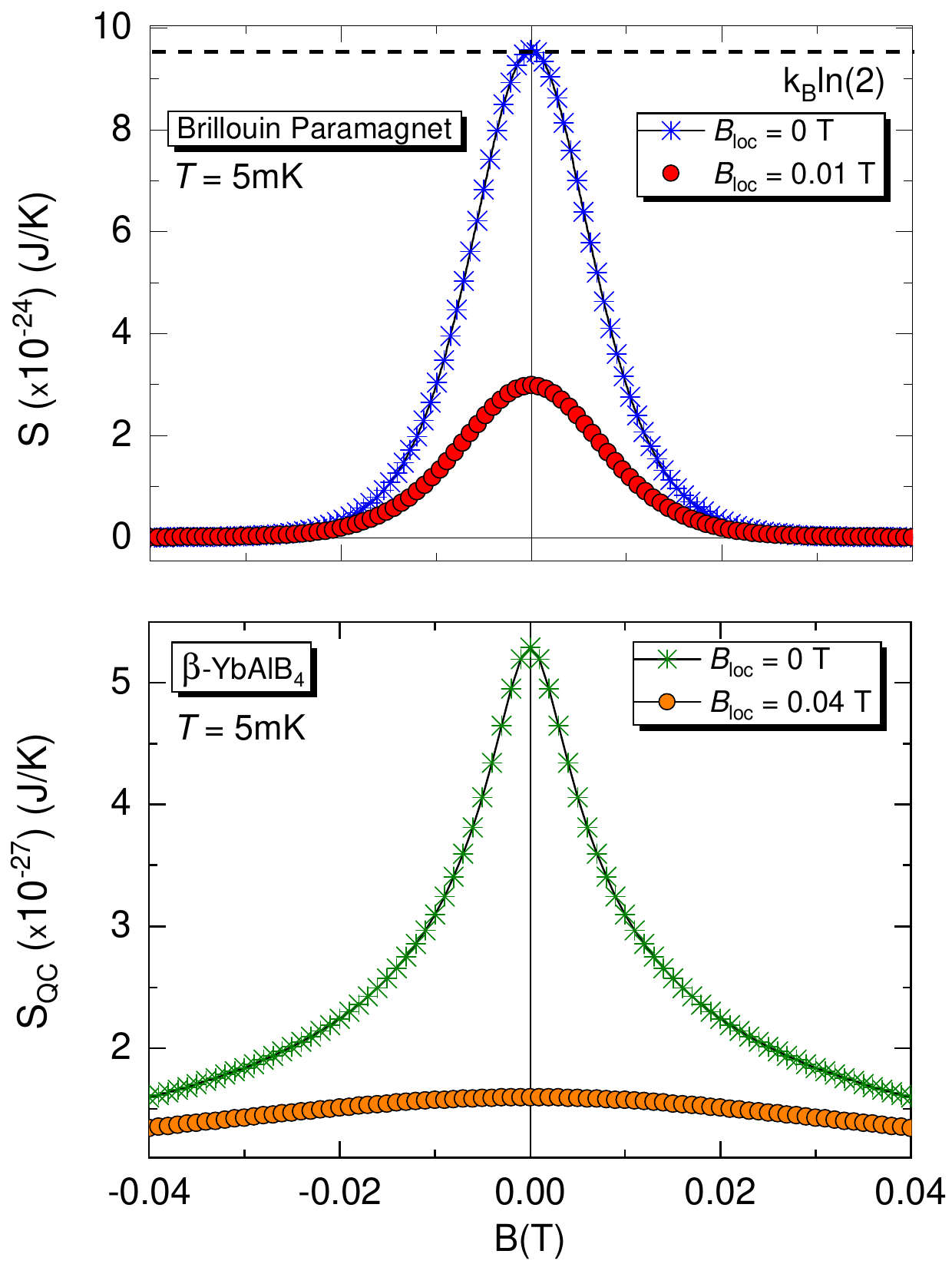}
\caption{\footnotesize \textbf{External magnetic field dependence of the entropy.} Entropy $S$ \emph{versus} external magnetic field $B$ for an arbitrarily fixed temperature of $T$ = 5\,mK. Upper panel: Brillouin paramagnet employing both $B_{loc}$ = 0 (blue asterisks) and 0.01\,T (red circles), considering neighboring magnetic moments separated by a distance $r =$ 5\,$\textmd{\AA}$. Lower panel: $S_{QC}$ \emph{versus} $B$ for $\beta$-YbAlB$_4$ employing $B_{loc}$ = 0 (green asterisks) and 0.04\,T (orange circles). Details in the main text.}
\label{Fig-2}
\end{figure}

We start our analysis based on entropy arguments.
In the case of an ideal paramagnet described by the textbook Brillouin-like model \cite{ralph}, the entropy can be derived from the probabilities of the spins (we consider $J = $ 1/2) to align parallel or anti-parallel to $B$ and it is given by:
\begin{equation}
S(B,T) = -\frac{B \mu_B}{T}\tanh{\left(\frac{B \mu_B}{k_B T}\right)} + k_B\textmd{ln}\left[2\cosh{\left(\frac{B \mu_B}{k_B T}\right)}\right],
\label{entropyBrillouin}
\end{equation}
where $k_B$ is the Boltzmann constant (1.38$\times$10$^{-23}$)\,J/K.
The entropy $S$ (Eq.\,\ref{entropyBrillouin}) as a function of $B$ at $T =$ 5\,mK (chosen arbitrarily), assuming $B_{loc}$ = 0  (blue asterisks) and $B_{loc} $ =  0.01\,T (red circles), is plotted in the upper panel of Fig.\,\ref{Fig-2}. Now, we recall the proposed quantum critical free-energy $F_{QC}$ for $\beta$-YbAlB$_4$ \cite{yosuke}:
\begin{equation}
F_{QC}(B,T) = -\frac{1}{(k_B \tilde{T})^{1/2}}{[(g\mu_B B)^2+(k_B T)^2]}^{3/4},
\label{F}
\end{equation}
where $k_B\tilde{T}$ $\approx$ 6.6\,eV $\approx$ (1.06$\times$10$^{-18}$)\,J, $\tilde{T}$ refers to a characteristic temperature and $g$ = 1.94 \cite{yosuke}. Using Eq.\,\ref{F}, it is straightforward to calculate the quantum critical entropy, $S_{QC}$ = $-(\partial F_{QC}/\partial T)_B$, namely:
\begin{equation}
S_{QC}(B,T) = \frac{3k_B^2 T}{2(k_B \tilde{T})^{1/2}(B^2 g^2{\mu_B}^2 + {k_B}^2 T^2)^{1/4}}.
\label{beta}
\end{equation}
Considering that $B_{loc}$ makes an angle $\theta \simeq$ 90$^{\circ}$ with $B$, we can write the resultant magnetic field  $B_r \simeq$ $\sqrt{B^2 + {B_{loc}}^2}$ \cite{ralph}, as depicted in the inset of Fig.\,\ref{Fig-1} b).
It turns out that when using $B_r$ directly in the partition function for the Brillouin paramagnet \cite{Reif} instead of $B$, the derived physical quantities will naturally have $B$ replaced by $B_r$ in their respective mathematical expressions. Thus, a key point of our analysis is that we have  replaced $B$ by $B_r$ in the expression of the entropy for the Brillouin paramagnet following the latter approach.

The obtained entropy $S_{QC}$ (Eq.\,\ref{beta}) as a function of $B$ for  $\beta$-YbAlB$_4$  considering arbitrarily chosen $T =$ 5\,mK, $B_{loc}$ = 0 (green asterisks) and $B_{loc} $  =  0.04\,T (orange circles), is plotted in the lower panel of Fig.\,\ref{Fig-2}. Essentially, in our analysis of the entropy we have used $B_r = B$ ($B_{loc}$ = 0\,T) and $B_r = \sqrt{B^2 + {B_{loc}}^2}$ in  Eqs.\,\ref{entropyBrillouin} and \ref{beta}.
As depicted in Fig.\,\ref{Fig-2}, for both the Brillouin-like paramagnet (upper panel) and $\beta$-YbAlB$_4$ (lower panel) the entropy is expressively lowered at zero external magnetic field when $B_{loc}$  is taken into account, since $B_{loc}$ favours long-range magnetic order \cite{ralph}.
Hence, the magnetic entropy is released and the third law of Thermodynamics is obeyed \cite{pathria}. These results suggest that, upon considering $B_{loc}$, the entropy accumulation when $B \rightarrow$ 0\,T is lowered and as a consequence $\Gamma_{mag}$ is only enhanced, but it does not diverge.

\begin{figure}[h!]
\centering
\includegraphics[scale=0.84]{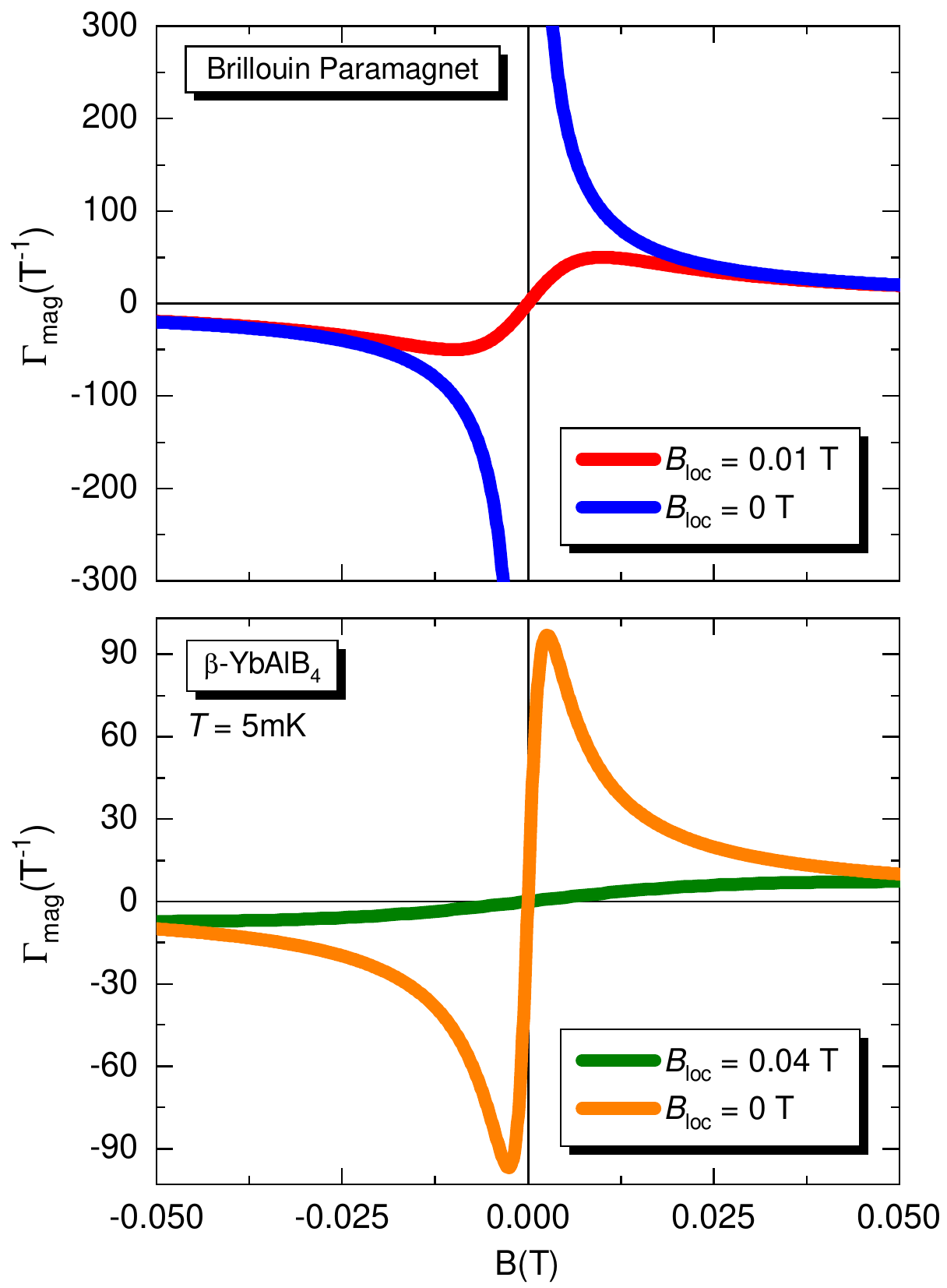}
\caption{\footnotesize \textbf{External magnetic field dependence of the magnetic Gr\"uneisen parameter.} Upper panel: Magnetic Gr\"uneisen parameter $\Gamma_{mag}$ \emph{versus} external magnetic field $B$  for the Brillouin paramagnet considering $B_{loc}$ = 0.01\,T (red line) and 0\,T (blue line), cf.\,labels.  Lower panel: $\Gamma_{mag}$ \emph{versus} $B$ for  $\beta$-YbAlB$_4$ considering  $B_{loc}$ = 0.04\,T (orange line) and 0\,T (green line) and $T$ = 5\,mK. Details in the main text.}
\label{Fig-4}
\end{figure}

\begin{figure}[h!]
\centering
\includegraphics[scale=0.73]{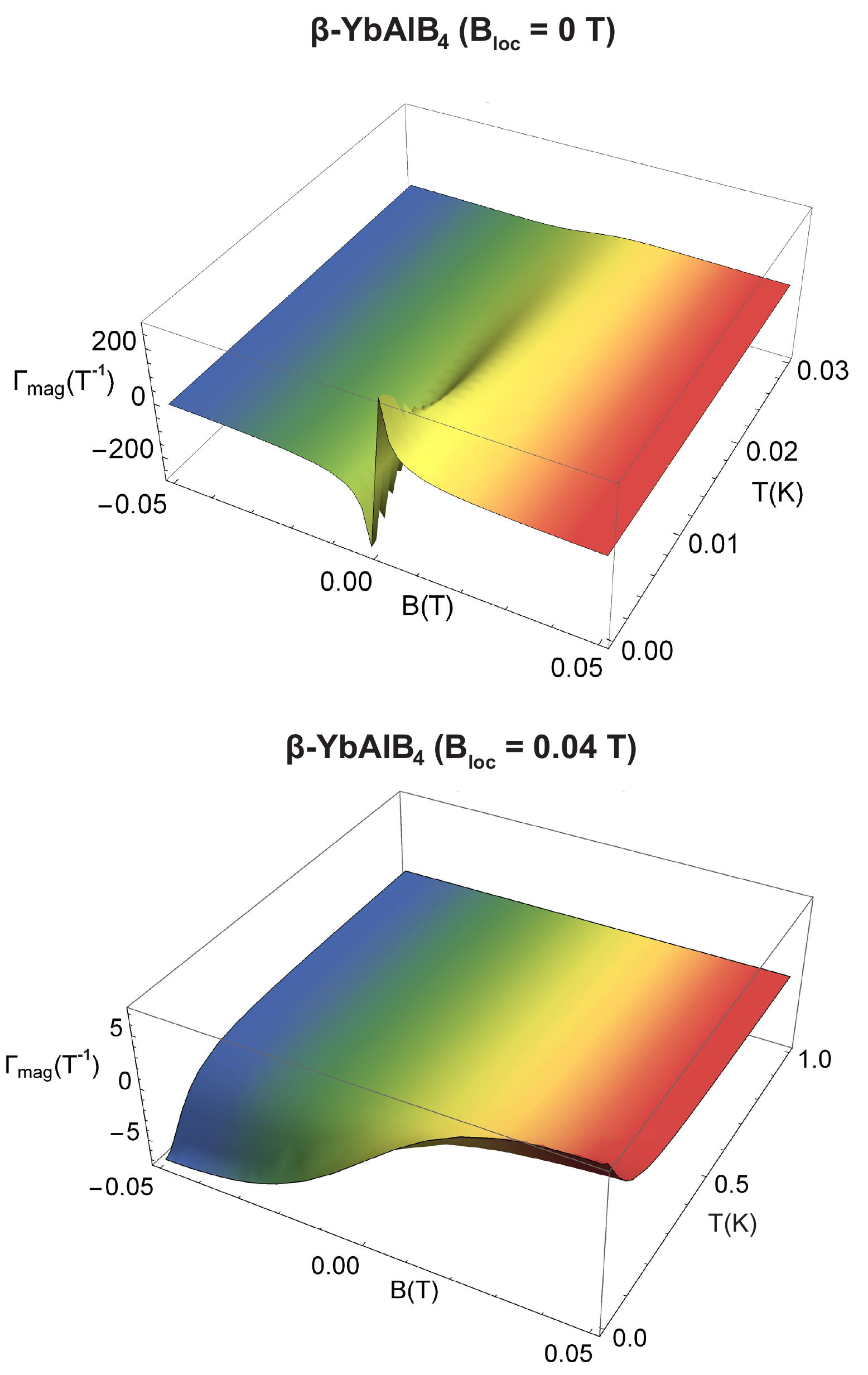}
\caption{\footnotesize  \textbf{External magnetic field and temperature dependence of the magnetic Gr\"uneisen parameter.} Magnetic Gr\"uneisen parameter $\Gamma_{mag}$ \emph{versus} external magnetic field $B$ \emph{versus} temperature $T$ for $\beta$-YbAlB$_4$ using $B_{loc}$ = 0 (upper panel) and $B_{loc} =$ 0.04\,T (lower panel). Note that $\Gamma_{mag}$  does not diverge in the limit of $T \rightarrow$ 0\,K and $B \rightarrow$ 0\,T when $B_{loc} \neq$ 0\,T. For clarity, two distinct temperature ranges were employed in the  plots. Further details in the main text.}
\label{Fig-3}
\end{figure}
We focus now on the analysis of $\Gamma_{mag}$, which can be calculated employing the well-known relation \cite{zhu}:
\begin{equation}
\Gamma_{mag} = -\frac{1}{T}\frac{\left(\frac{\partial S}{\partial B}\right)_T}{\left(\frac{\partial S}{\partial T}\right)_B},
\label{Gamma}
\end{equation}
taking into account $B_{loc}$. Hence, in the following we consider in addition to the external magnetic field ($B$) the effects of $B_{loc}$ on $\Gamma_{mag}$. We thus replace $B$ by $B_r$ in Eqs.\,\ref{entropyBrillouin} and \ref{beta}, so that:
\begin{equation}
\Gamma_{mag} = \frac{B}{B^2 + {B_{loc}}^2},
\label{gammabri2}
\end{equation}
for the Brillouin paramagnet \cite{prbmce}, while for $\beta$-YbAlB$_4$, $\Gamma_{mag}$ reads:
\begin{equation}
\Gamma_{mag} = \frac{B g^2{\mu_B}^2}{{k_B}^2T^2 + 2g^2{\mu_B}^2(B^2+B_{loc}^2)}.
\label{gammapiers}
\end{equation}
The results of $\Gamma_{mag}$ for the Brillouin paramagnet and $\beta$-YbAlB$_4$ (Eqs.\,\ref{gammabri2} and \ref{gammapiers}, respectively) are shown in Figs.\,\ref{Fig-4} and \ref{Fig-3}.
A careful analysis of  $\Gamma_{mag}$ for the Brillouin paramagnet (upper panel of Fig.\,\ref{Fig-4}) enables us to relate our findings with the seminal experiment proposed by Purcell and Pound \cite{Purcell} regarding the achievement of negative temperatures in laboratory \cite{ralph}, as well as with the well-known impossibility of achieving absolute zero temperature for $B_{loc} \neq$ 0\,T, to be discussed in the next subsection. In the case of $\beta$-YbAlB$_4$, $\Gamma_{mag} \rightarrow \infty$  for  $T \rightarrow$ 0\,K and $B \rightarrow$ 0\,T when $B_{loc}$ = 0\,T (upper panel of Fig.\,\ref{Fig-3}), but it does not do so for $B_{loc} \neq$ 0\,T (lower panel of Fig.\,\ref{Fig-3}).
Hence, the  consequence of taking $B_{loc} \neq$ 0\,T into account is that $\Gamma_{mag}$ will never diverge for $T\rightarrow$ 0\,K and $B \rightarrow$ 0\,T, since $B_r$ will never be zero and thus we cannot infer a genuine zero-field quantum critical point. This is one of the main results of our work. We stress that we are not dealing with a simple shift in the position of the maximum value of $\Gamma_{mag}$, cf.\,Eq.\,\ref{gammabri2}. The situation is quite different, for instance, for the one-dimensional Ising model under transverse magnetic field \cite{Si}, where the critical field for the divergence of $\Gamma_{mag}$ is shifted when the ratio of the critical field to coupling constant between nearest neighbor is changed \cite{Si}.
When analyzing Eqs.\,\ref{gammabri2} and \ref{gammapiers}, respectively, for the Brillouin paramagnet and $\beta$-YbAlB$_4$, considering $B_{loc}$ $\neq$ 0\,T, we observe that the maximum of $\Gamma_{mag}$ is centered at $B$ = $B_{loc}$ for the Brillouin paramagnet, while for $\beta$-YbAlB$_4$ it is centered at  $B$ = $[{B_{loc}}^2$ + $({k_B}^2 T^2)/(2g^2 {\mu_B}^2)]^{1/2}$, i.e., when $T$ = 0\,K the maximum is located at $B$ = $B_{loc}$, cf.\,Fig.\,\ref{Fig-4}. The maxima were obtained by the simple optimization of the $\Gamma_{mag}$ functions (Eqs.\,\ref{gammabri2} and \ref{gammapiers}), i.e., making $\left(\partial\Gamma_{mag}/\partial B\right)_T$ = 0 in both cases.
Yet, it is worth mentioning that the position of the maximum value of  $\Gamma_{mag}$ is related to the well-known Schottky anomaly \cite{blundell,pathria}, to be discussed into more details in the following.
For $\beta$-YbAlB$_4$, under the condition $k_B T  = \mu_B B_{loc}$, considering $B_{loc}$ = 0.04\,T, we obtain $T = \mu_B B_{loc}/k_B \simeq$ 24\,mK. The latter indicates the temperature onset of $B_{loc}$ relevance  for this system. It is worth mentioning that the lowest temperature of the experiments reported in Fig.\,2(A) of Ref.\,\cite{yosuke} for $\beta$-YbAlB$_4$ was $T \simeq$ 10\,mK and the lowest external magnetic field was 0.31\,mT. In terms of Maxwell-relations, namely $\left(\frac{\partial M}{\partial T}\right)_B = \left(\frac{\partial S}{\partial B}\right)_T$, being $M$ the magnetization, it is tempting to say that the results of Fig.\,2(A) of Ref.\,\cite{yosuke} are at odds with a diverging $\Gamma_{mag}$.
Yet, considering $\beta$-YbAlB$_4$, our analysis is corroborated by the results presented in Fig.\,S2 of Ref.\cite{yosuke}, namely  $\Gamma_{mag}$ \emph{versus} external magnetic field for various temperatures. There, a  clear decrease of $\Gamma_{mag}$ was observed experimentally upon decreasing $B$. However, in the frame of zero-field quantum criticality, we would expect an enhancement of $\Gamma_{mag}$ for $T \rightarrow$ 0\,K and $B \rightarrow$ 0\,T \cite{gegenwart}.  In Ref.\cite{Matsukawa}, the authors discuss that the free energy scaling for the systems CeCu$_{6-x}$Au$_x$ and Au-Al-Yb, approximant and quasicrystal, respectively, follow the same scaling behavior as  $\beta$-YbAlB$_4$. These systems are also considered to be zero-field quantum critical. Then, our analysis can be extended to \emph{all} systems that follow the same scaling behavior of the free energy reported in Ref.\,\cite{yosuke}, see Eq.\,\ref{F}. The main result of this subsection is that zero-field quantum criticality will not hold in \emph{any} system where resultant magnetic moments are non-negligible.
A corresponding situation is also observed for the one-dimensional (1D) Ising model under longitudinal magnetic field \cite{prbmce,luquinha}, where the magnetic coupling constant $J'$ (we make use of $J'$ to avoid confusion with the momentum quantum number $J$) plays the role analogously to $B_{loc}$, i.e.,  $\Gamma_{mag}$ only diverges for $J' \rightarrow$ 0. In fact, the 1D Ising model under longitudinal field \cite{prbmce} is equivalent to the Brillouin paramagnet in two distinct cases, namely: \emph{i}) at the ferromagnetic ground state where $J'$ can be associated with a local magnetic field, which in turn acts as $B_{loc}$ and; \emph{ii}) in the limit $T \rightarrow \infty$ for finite $B$, where due to the increase of the thermal energy the ferromagnetic ordering is suppressed giving rise to a paramagnetic phase \cite{prbmce}. Hence, considering the similar role played by $J'$ and $B_{loc}$, it becomes evident that $\Gamma_{mag}$ only diverges for vanishing values of $J'$, analogously to the case for the Brillouin paramagnet when $B_{loc}$ = 0\,T, as discussed previously. In general terms, \emph{any} system with finite mutual interactions will not show a diverging $\Gamma_{mag}$ and thus zero field quantum criticality cannot take place. This analysis reinforces the universal character of the mutual interactions and their role in the field of quantum criticality.
In the frame of the original work reported by Weiss \cite{weiss}, the molecular field is associated with the mutual interactions, which in turn leads to the ordering of the magnetic moments within the regime of relevance of such interactions \cite{kittel}. Hence, in the same way as the Curie temperature \cite{blundell} represents the critical temperature for ferromagnets, analogously for the mutual interactions we can infer a pseudo \textit{critical temperature} $T_c$, which defines their regime of relevance.

\vspace{0.3cm}
\textbf{\emph{ii}) Connection between the canonical definition of temperature and $\Gamma_{mag}$.}

\begin{figure}[h!]
\centering
\includegraphics[scale=1]{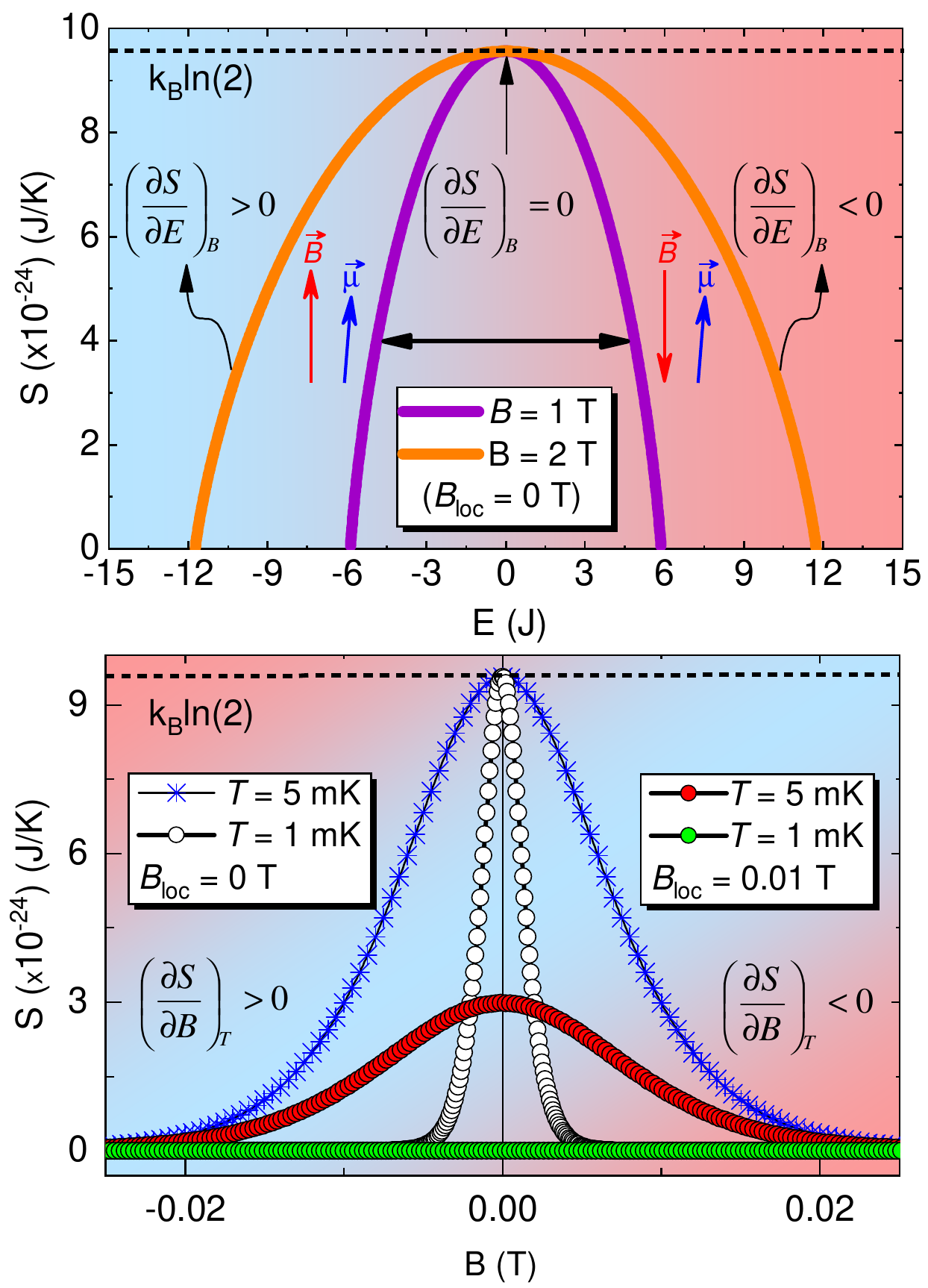}
\caption{\footnotesize \textbf{Average magnetic energy and external magnetic field dependence of the entropy.} Upper panel: Entropy $S$ as a function of the average magnetic energy $E$ for $B$ = 1,  2\,T; $B_{loc}$ = 0\,T and $N = (6.022 \times 10^{23})$.
The horizontal black arrow indicates the possibility of changing the sign of the magnetic energy when the direction of the external magnetic field (red arrow) is varied in the time scale of $\mu$s resulting predominantly in an anti-parallel configuration of the magnetic moment in respect to $\vec{\mu}$ (blue arrow), as proposed in the Purcell and Pound's experiment \cite{Purcell}. The colorful background indicates that negative (red) temperatures are hotter than positive (blue) ones. Lower panel: Entropy $S$ as a function of the external magnetic field $B$ for the Brillouin model at $T$ = 5 and 1\,mK considering $B_{loc}$ = 0\,T and $B_{loc}$ = 0.01\,T, cf.\,label. The dashed lines in both panels represent the entropy saturation $S = k_B\ln$(2), i.e., the entropy value for $(\mu_B B/k_B T)  \rightarrow$ 0. The colorful background depicts the insensitivity of $(\partial S/\partial B)_T$ regarding positive or negative temperatures, cf.\,discussed in the main text.}
\label{Fig-5}
\end{figure}

Considering the definition of temperature 1/$T$ = $\left(\partial S/\partial E\right)_B$ \cite{ralph,pathria}, absolute zero temperature can be inferred when $\left(\partial S/\partial E\right)_B \rightarrow \infty$ (upper panel of Fig.\,\ref{Fig-5}), where $E$ refers to the average magnetic energy, being $E = k_B T^2 \frac{\partial \ln Z}{\partial T} = -\mu_B B N\tanh{\left(\frac{\mu_B B}{k_B T}\right)}$, with $Z = 2\cosh{\left(\frac{\mu_B B}{k_B T}\right)}$ the partition function for the Brillouin paramagnet \cite{ralph} and $N$ refers to the number of particles. Upon analyzing the behavior of the entropy $S$ as a function of the external magnetic field at various temperatures for the Brillouin paramagnet, depicted in the lower panel of Fig.\,\ref{Fig-5}, we observe that there is an intrinsic entropy accumulation when $B \rightarrow $ 0\,T. This is naturally expected, since for $B =$ 0\,T the entropy achieves its maximum value, namely $S = k_Bln(2)$.
From the expression for the average magnetic energy, we can easily write $T(E,B)$ as:
\begin{equation}
T(E,B) = \frac{\mu_B B}{k_B \textmd{arctanh}\left(-\frac{E}{\mu_B B N}\right)},
\end{equation}
and then plug it in the expression for the entropy (Eq.\,\ref{entropyBrillouin}) obtaining thus $S(E,B)$:
\begin{equation}
S(E,B) = -\frac{Ek_B \textmd{arctanh}\left(\frac{E}{\mu_B B N}\right)}{B \mu_B N} + k_B\ln{\frac{2}{\sqrt{1 - \frac{E^2}{B^2 {\mu_B}^2 N^2}}}} \label{S-3}.
\end{equation}
Note that when $E =$ 0\,J in Eq.\,\ref{S-3}, $S = k_Bln(2)$ is nicely recovered, cf.\,upper panel of Fig.\,\ref{Fig-5}.
Then, employing Eq.\,\ref{S-3}, we compute separately the magnetic field and energy derivatives of $S$, namely $\left(\frac{\partial S}{\partial B}\right)_T$ and $\left(\frac{\partial S}{\partial E}\right)_B$, respectively. It turns out that both derivatives are related to each other by the simple form:
\begin{equation}
\left(\frac{\partial S}{\partial B}\right)_T = \frac{E}{B}\left(-\frac{\partial S}{\partial E}\right)_B \label{S-4}.
\end{equation}
Based on Eq.\,\ref{S-4} we discuss in the next the connection between ($\partial S$/$\partial E$)$_B$ and ($\partial S$/$\partial B$)$_T$.
Employing the expression for the average magnetic energy, discussed previously, and upon analyzing Fig.\,\ref{Fig-5} (upper panel), it is clear that negative values of $E$ correspond to positive values of $T$, since the employed values of $B$ = 1 and 2\,T  in our calculations are positive and constant making thus $T$ the quantity that rules the sign of $E$. Hence, upon analyzing the results shown in the lower panel of Fig.\,\ref{Fig-5}, we can infer directly neither negative, infinite nor absolute zero temperature in the same way as upon analyzing the upper panel of Fig.\,\ref{Fig-5}, since a fixed positive value of $T$ was employed. However, for finite values of $B$  when the limit $T \rightarrow$ $\infty$ is considered in the lower panel of Fig.\,\ref{Fig-5}, the entropy saturates ($S$ = $k_B$$\ln$2) and thus $\left(\partial S/\partial B\right)_T \rightarrow$ 0. In the same way, when $T \rightarrow$ 0\,K, $\left(\partial S/\partial B\right)_T \rightarrow$ $\infty$. This is in agreement with both a diverging (zero temperature) and vanishing (infinite temperature) of $\left(\partial S/\partial E\right)_B$. However, strictly speaking $\left(\partial S/\partial E\right)_B$ will never diverge since $\vec{\mu}$ and $\vec{B}$ cannot be perfectly aligned, as discussed previously, and thus the achievement of absolute zero temperature is prevented. Yet, when the local magnetic field  $B_{loc}$ = 0\,T (blue asterisks and open circles in the lower panel of Fig.\,\ref{Fig-5}), upon approaching $T$ = 0\,K, $\left(\partial S/\partial B\right)_T \rightarrow \infty$ and thus absolute zero temperature can be indirectly inferred. However, when $B_{loc} \neq$ 0\,T (red and green circles in the lower panel of Fig.\,\ref{Fig-5}), for vanishing $T$ and $B$, such an entropy accumulation [$\left(\partial S/\partial B\right)_T \rightarrow$ $\infty$] is suppressed and, as a consequence, the achievement of absolute zero temperature is also prevented. Evidently, such a divergence of the entropy for vanishing $T$ and $B$ would violate the third law of Thermodynamics \cite{ashcroft}, strengthening thus our argument that a genuine zero-field QCP cannot take place, cf.\,discussions in the previous subsection. Thus, we have demonstrated in two different ways that the achievement of absolute zero temperature is not possible when considering finite $B_{loc}$.
Also, in the lower panel of Fig.\,\ref{Fig-5}, when $B_{loc}$ = 0\,T, the entropy for $B =$ 0\,T is exactly the same for both data set (blue asterisks and open circles), since for $B$ = 0\,T the resultant entropy $S$ = $k_B \ln(2)$ is nicely recovered for any $T \neq$ 0\,K \cite{ashcroft}. This is not the case when $B_{loc}$ is taken into account, being that for $B$ = 0\,T the entropy  depends on $T$ \cite{ashcroft}, as depicted in the lower panel of Fig.\,\ref{Fig-5}.

\begin{figure}[h!]
\centering
\includegraphics[width=\textwidth]{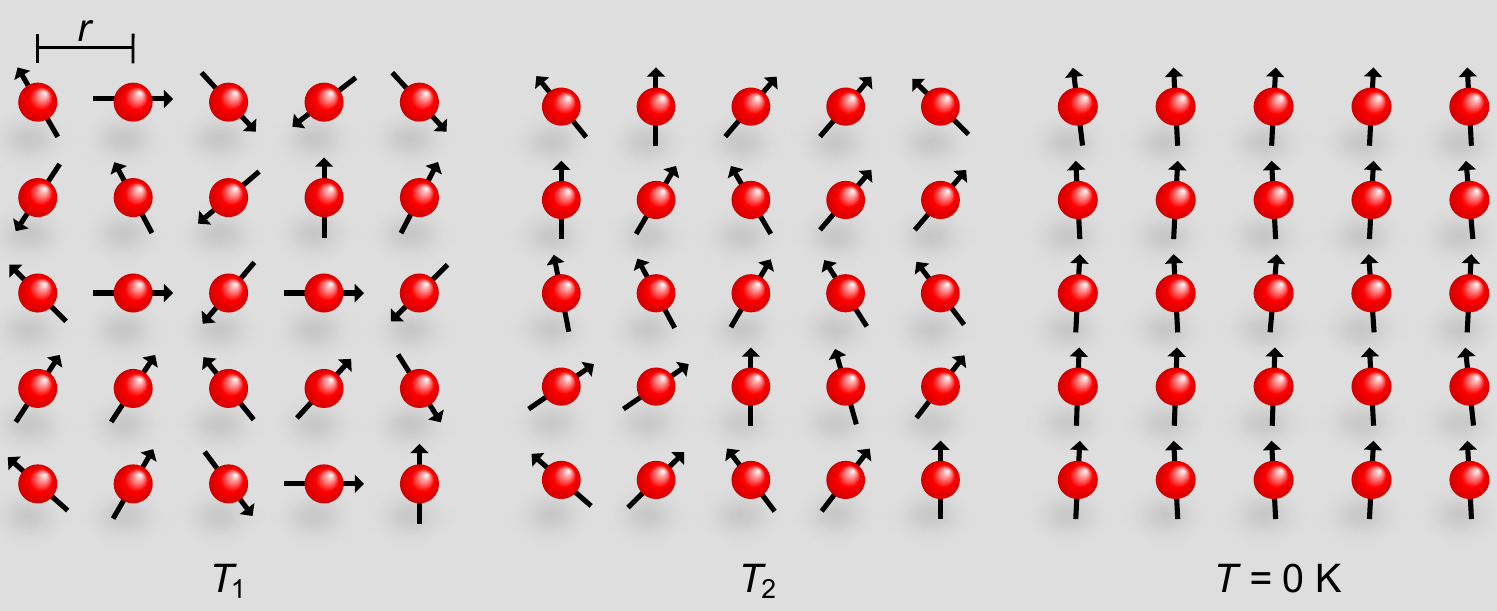}
\caption{\footnotesize \textbf{The ground-state of a paramagnet.} Schematic representation of the intrinsic magnetic moments (black arrows) of the electrons (red spheres), separated by a distance $r$, for the Brillouin paramagnet. Upon lowering the temperature $T_1$ $>$ $T_2$ $>$ $T$ = 0\,K in the absence of external magnetic field, all magnetic moments are naturally aligned along the same direction and a magnetically ordered phase takes place, since an effective local magnetic field $B_{loc}$ is intrinsically present due to the mutual interactions between adjacent magnetic moments \cite{kittel, ashcroft, ralph}. Note the lack of perfect alignment of the magnetic moments in respect to $B_{loc}$, cf.\,discussion in the main text.}
\label{Fig-6}
\end{figure}
The entropy as a function of the average energy (upper panel of Fig.\,\ref{Fig-5}) can be easily obtained through Eq.\,\ref{S-3} employing the arbitrarily fixed values of $B$ = 1 and 2\,T. Figure \ref{Fig-6} depicts schematically the relevance of $B_{loc}$ for $T \rightarrow$ 0\,K and $B \rightarrow$ 0\,T, since a ferromagnetic ground state only emerges due to the presence of finite mutual interactions in the system. Recalling that the absence of long-range magnetic ordering in low-dimensional systems at finite temperature  is a direct consequence of the Mermin-Wagner theorem \cite{Mermin}.   Interestingly enough, when considering the equilibrium spin populations in a two-level system, well known from classical textbooks \cite{kittel, ashcroft, ralph}, namely:

\begin{equation}
\frac{N_1}{N_T} = \frac{\exp(\mu B/k_B T)}{\exp(\mu B/k_B T)\,+\,\exp(-\mu B/k_B T)}\label{popu1}
\end{equation}
and
\begin{equation}
\frac{N_2}{N_T} = \frac{\exp(-\mu B/k_B T)}{\exp(\mu B/k_B T)\,+\,\exp(-\mu B/k_B T)},\label{popu2}
\end{equation}

being $N_1$ and $N_2$ the spin populations regarding the lower and upper energy levels, respectively, and $N_T$ the total number of spins. It turns out that for vanishing $T$ and positive values of $B$, the ratio $\frac{N_1}{N_T}$ becomes 1 and $\frac{N_2}{N_T}$ is zero, which means that all spins point roughly along the same direction, i.e., a ferromagnetic ordering takes place, as schematically shown in Fig.\,\ref{Fig-6} for $T$ = 0\,K. In the following we make a link between $\Gamma_{mag}$, the definition of temperature and the Purcell and Pound's experiment \cite{Purcell}. Before starting the discussions, we stress that we are not proposing a new definition of temperature.
Recalling that the average magnetic energy is given by \cite{ralph}:
\begin{equation}
E = -N\mu_B B\tanh{\left(\frac{\mu_B B}{k_BT}\right)},
\end{equation}
it is possible to rewrite $E$ as a function of the average magnetic moment  $\langle\mu\rangle = N \mu_B \tanh{\left(\mu_B B/ k_B T\right)}$ along the external magnetic field,  as follows:
\begin{equation}
E = -\langle\mu\rangle B.
\end{equation}
From the expression for $\langle\mu\rangle$ it is possible to write the external magnetic field in respect to $\langle\mu\rangle$:
\begin{equation}
B(T, {\langle\mu\rangle}) = \frac{k_B T}{\mu_B} \textmd{arctanh}\left(\frac{\langle\mu\rangle}{N\mu_B}\right).
\label{averagemagneticmoment}
\end{equation}
Equation \ref{averagemagneticmoment} indicates that at a certain temperature $T$, the value of an external magnetic field $B$ is associated with a spin configuration that corresponds to a specific average magnetic moment $\langle\mu\rangle$. Recalling that the magneto-caloric effect can be quantified by $\Gamma_{mag}$ \cite{zhu}:
\begin{equation}
\Gamma_{mag}=\frac{1}{T}\left(\frac{\partial T}{\partial B}\right)_S,
\label{MCE-Eq}
\end{equation}
we can compute the temperature derivative of $B$ in Eq.\,\ref{averagemagneticmoment} and replace it into Eq.\,\ref{MCE-Eq}, resulting:
\begin{equation}
\Gamma_{mag} = \frac{\mu_B}{Tk_B\textmd{arctanh}\left(\frac{\langle\mu\rangle}{N\mu_B}\right)}.
\end{equation}
Since $\Gamma_{mag}$ = 1/$B$ for the Brillouin paramagnet \cite{prbmce}, we write $T$ as a function of $B$ and $\langle\mu\rangle$:
\begin{equation}
T = \frac{B \mu_B}{k_B \textmd{arctanh}\left(\frac{\langle\mu\rangle}{N\mu_B}\right)}.
\label{connection}
\end{equation}
Equation \ref{connection} connects in an unprecedent way the Purcell and Pound experiment \cite{Purcell} and $\Gamma_{mag}$ itself. When $B$ and $\langle\mu\rangle$ are $\parallel$, positive temperatures are inferred. However, when $B$ and $\langle\mu\rangle$ are anti-$\parallel$, then negative temperatures can be associated.
\begin{figure}[h!]
\centering
\includegraphics[scale=1.00]{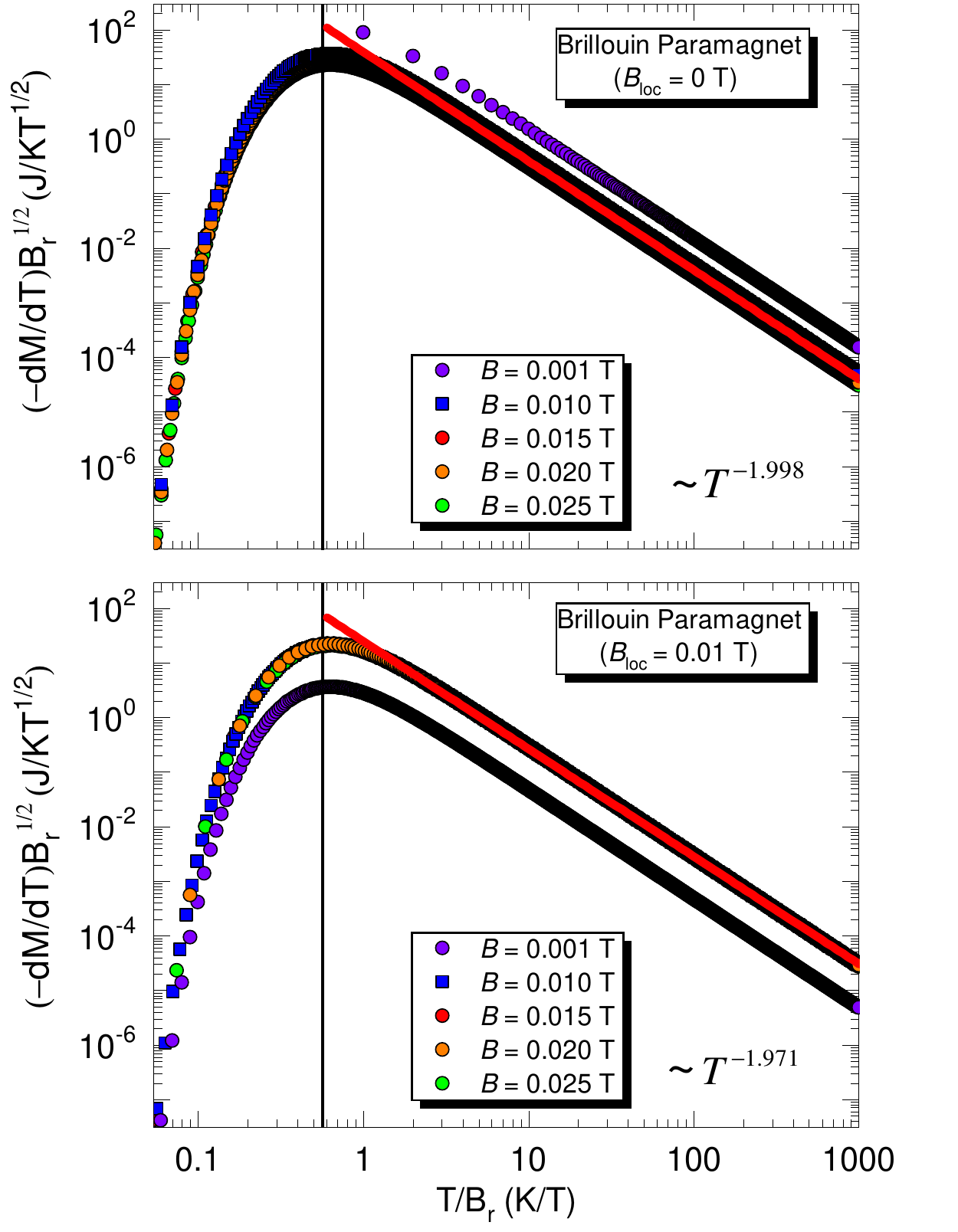}
\caption{\footnotesize \textbf{Scaling analysis for the Brillouin paramagnet.} Scaling behavior of the magnetization for the Brillouin paramagnet for fixed values of external magnetic field considering $B_{loc}$ = 0\,T (upper panel) and $B_{loc}$ = 0.01\,T (lower panel).
The solid vertical lines in both panels indicate the position of the Schottky maximum, while the red solid line indicates the overlap (scaling) of all data set in the regime of high temperatures.
The scaling exponent is $\epsilon$ = 1, the same one of $\beta$-YbAlB$_4$, and d$M$/d$T$ scales with temperature as $\sim T^{-1.998}$ (upper panel) and as $\sim T^{-1.971}$ (lower panel).
Details in the main text.}
\label{Fig-7}
\end{figure}
It is remarkable that the definition of temperature is encoded in $\Gamma_{mag}$ and vice-versa. Next, we focus on the scaling analysis of the magnetization. When investigating the behavior of $\Gamma_{mag}$, an important consideration is its scaling behavior, see, e.g.\,Ref.\,\cite{gegenwart} and references cited therein.
Upon analyzing the scaling behavior for the Brillouin paramagnet, shown in Fig.\,\ref{Fig-7}, we observe that all data collapse in the same line only for external magnetic field values close to $B_{loc}$. By considering much higher or lower values of external magnetic field when compared with $B_{loc}$, a breakdown of the scaling behavior of the magnetization is observed. As a matter of fact, the scaling analysis reflects the behavior of $\Gamma_{mag}$ as a function of $T$, so that the maxima depicted in Fig.\,\ref{Fig-7} are associated with the Schottky anomaly, which in turn is only observed in the presence of mutual interactions.
In other words, due to the intrinsic mutual interactions present into the system, the validity of the scaling in the Brillouin paramagnet is limited only to external magnetic field values close to $B_{loc}$. Thus, our scaling analysis also demonstrates the relevance of $B_{loc}$ in the limit of both $B \rightarrow$ 0\,T and $T \rightarrow $ 0\,K. In Fig.\,S3 of Ref.\,\cite{yosuke} the authors show a scaling behavior of $\beta$-YbAlB$_4$ upon analyzing the critical contribution of the magnetization $M_c$. Interestingly enough, there is a breakdown of the scaling behavior for applied magnetic fields higher than 0.5\,T, which resembles the scaling behavior of the Brillouin paramagnet discussed here, and the scaling is only valid for low values of $B$ for $\beta$-YbAlB$_4$.
At this point, it is worth recalling that $\Gamma_{mag}$ can also be written as follows \cite{zhu}:
\begin{equation}
\Gamma_{mag} = - \frac{(dM/dT)_B}{c_B},
\label{dmdt}
\end{equation}
where $c_B$ is the specific heat at constant external magnetic field.
Upon employing Eq.\,\ref{dmdt}, it can be directly inferred that when $\left(- dM/dT\right)_B \rightarrow \infty \Rightarrow \Gamma_{mag} \rightarrow \infty$ if $c_B$ is non-singular. Thus, we make use of Eq.\,\ref{dmdt} in order to infer the divergence of $\Gamma_{mag}$ for the scaling analysis of the Brillouin paramagnet, as shown in Fig.\,\ref{Fig-7}. It is worth mentioning that since $\Gamma_{mag}$ for the Brillouin paramagnet is temperature independent \cite{prbmce}, we thus make use of the magnetization scaling in our analysis. Interestingly, also for the one-dimensional Ising model under transverse magnetic field  $\Gamma_{mag}$ also independs on $T$ \cite{Si}.
We observe from Fig.\,\ref{Fig-7} that when $B_{loc}$ = 0\,T,  for arbitrarily chosen $B$ = 0.001\,T,   the scaling consists in a straight line. Since a logarithm scale was used in both axes of Fig.\,\ref{Fig-7}, the straight line indicates a divergence of $\Gamma_{mag}$ for  $B \rightarrow$ 0\,T and $T \rightarrow$ 0\,K. However, when $B_{loc}$ = 0.01\,T the divergence of $\Gamma_{mag}$ is suppressed and such linear behavior is no longer observed, opening the way for the appearance of a Schottky-like maximum. Hence, the scaling plots depicted in Fig.\,\ref{Fig-7} demonstrate in another way the non-divergence of $\Gamma_{mag}$ when $B_{loc} \neq$ 0\,T for the Brillouin paramagnet.
Also, the maximum in $\Gamma_{mag}$ shown in both panels of Fig.\,\ref{Fig-4} when $B_{loc} \neq $ 0\,T  is due to the Schottky anomaly \cite{blundell,pathria}, as previously mentioned. As well-known from textbooks, such maximum takes place when $(k_B T)/(\mu_B B) \simeq $ 0.834 \cite{blundell,tari}. Upon continuously decreasing the temperature of the system, the spins occupy preferably the lower energy level and, as a consequence, all spins will occupy the same energy level in the ground state. This can be visualized in the spins scheme depicted in Fig.\,\ref{Fig-6}, and it is in line with our analysis for the spin populations, cf.\,Eqs.\,\ref{popu1} and \ref{popu2}. The Schottky anomaly can also be captured in the scaling analysis for the Brillouin model (Fig.\,\ref{Fig-7}) where a maximum takes place at $T/B_r$ $\approx$ 0.5602. Now, we use again the definition of the magneto-caloric effect (Eq.\,\ref{MCE-Eq}) \cite{zhu}. Knowing that $\Gamma_{mag} = 1/B$ for the Brillouin paramagnet \cite{prbmce}, it is straightforward to write $\frac{T}{B} = \left(\frac{\partial T}{\partial B}\right)_S$. Hence, this simple analysis indicates that temperature and magnetic field are interconnected. Indeed, in the adiabatic demagnetization the temperature is decreased upon removing the external applied magnetic field, in order to keep the ratio $(\mu_B B_r)/(k_B T)$ constant.

\vspace{0.3cm}
\textbf{\emph{iii}) Possibility of performing adiabatic magnetization by only manipulating the mutual interactions.}

Our proposal is distinct from the adiabatic demagnetization method itself, since no external magnetic field is required.
As pointed out in Ref.\,\cite{ralph}, although counter-intuitive it is possible to increase the temperature adiabatically. The idea behind is based on the connection between the uncertainty principle and the entropy. The entropy of the system can be written as follows \cite{ralph}:
\begin{equation}
S = -k_B \sum_{j} P(\Psi_j)\ln(\Psi_j),
\label{psi}
\end{equation}
where $P(\Psi_j)$ is the probability of the system to be at the energy Eigenstate $\Psi_j$ and $j$ is the label of the corresponding Eigenstate.
\begin{figure}[h!]
\centering
\includegraphics[scale=0.60]{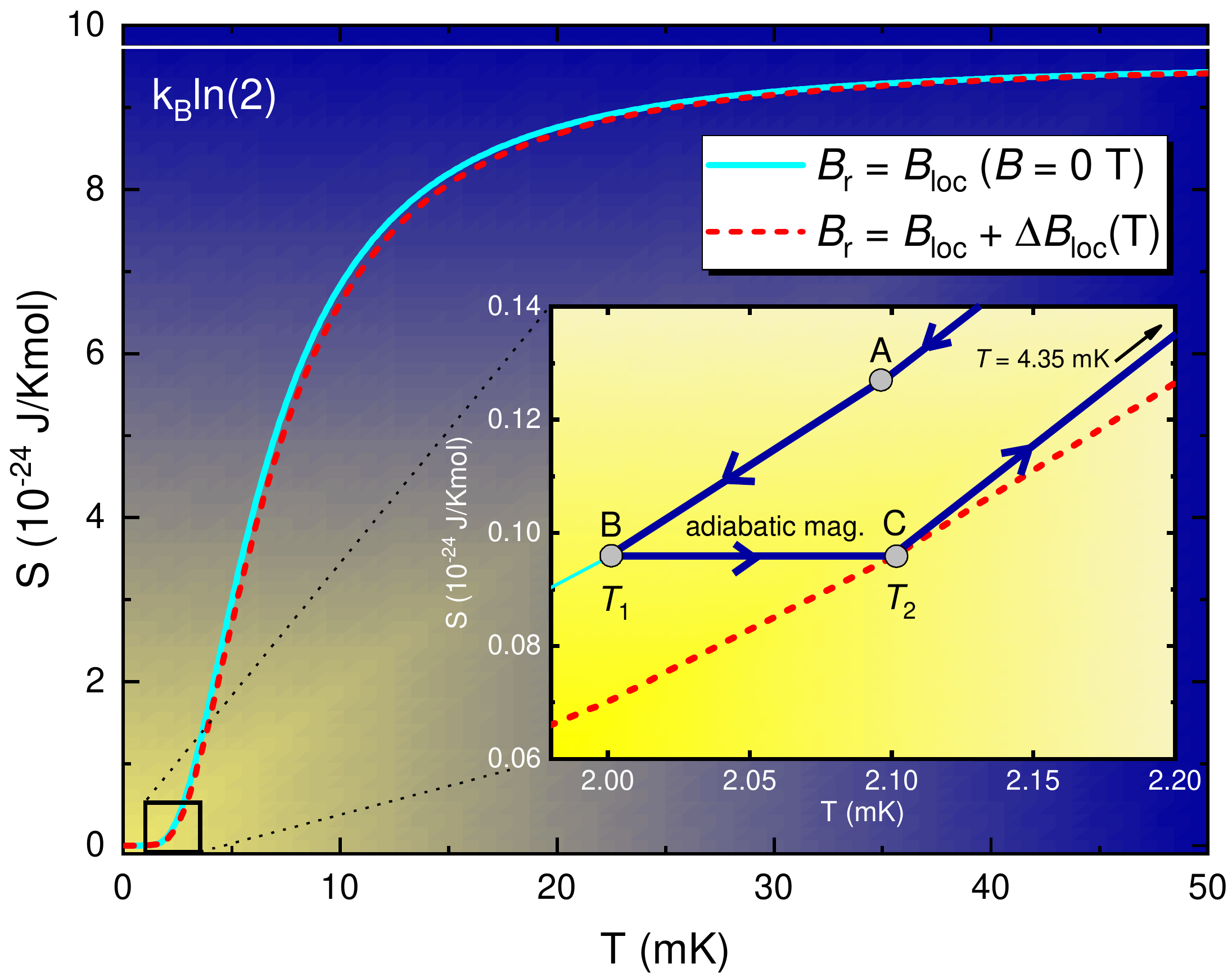}
\caption{\footnotesize \textbf{Adiabatic demagnetization of the mutual interactions.} Main panel: entropy as a function of temperature for $B = 0$\,T (solid cyan curve) and for $\Delta B_{loc}$ = 0.0032\,T (red dashed line). The yellow gradient indicates the region of relevance of $B_{loc}$ in the regime of ultra-low temperatures. Inset: schematic representation of the adiabatic magnetization employing the mutual interactions. The steps are as follows: \emph{i}) the system is cooled down to a sufficiently low temperature where the magnetic energy associated with $B_{loc}$ overcomes the thermal fluctuations (path A to B); \emph{ii}) the temperature is adiabatically increased from B ($T_1$) to C ($T_2$) and, in order to hold the entropy constant, the system is magnetized adiabatically, i.e., a finite value of $\Delta B_{loc}$ emerges into the system and thus, the entropy changes from the solid cyan to the \emph{fictitious} red dashed line, which emulates $S$ \emph{versus} $T$ for the corresponding value of $(B_{loc} + \Delta B_{loc})$; \emph{iii}) upon increasing the temperature from C, the entropy will vary and, as a consequence, the system will be demagnetized at $T$ = 4.35\,mK (indicated by the solid black arrow), going back to its original configuration of $\Delta B_{loc}$ = 0\,T. Such process can be restarted by cooling the system again from $T$ = 4.35\,mK to point A.}
\label{Fig-8}
\end{figure}
It is clear that the entropy increases with the uncertainty associated with the particle according with Eq.\,\ref{psi} and thus, in order to hold the entropy constant, the uncertainty must also be held constant.
In an adiabatic expansion, the increase of the volume causes an enhancement in the spatial uncertainty \cite{ralph}. However, in this process, the gas particles that do work will lose energy and then they will have a decrease in their linear momentum, i.e., the momentum uncertainty is lowered as well. Thus, there is a compensation of such uncertainties and, as a consequence, the entropy remains constant during this process \cite{ralph}. Following similar arguments, adiabatic magnetization of the mutual interactions can be achieved. As depicted in Fig.\,\ref{Fig-8}, upon increasing the temperature adiabatically, a spontaneous magnetization takes place as a direct consequence of the constrain of holding the entropy constant. Since there is no external applied magnetic field we have $B_r = B_{loc}$ and, in order to hold $S$ (Eq.\,\ref{entropyBrillouin}) constant, the magnetic energy should also be changed in this process, so that adiabatic magnetization is achieved. This can also be easily understood in terms of the uncertainty principle.  It is straightforward to calculate the uncertainty $\Delta E = U_{mag} - E$ of the magnetic energy, which reads:
 \begin{equation}
 \Delta E = \mu_B B_{loc}\left[N \tanh \left(\frac{\mu_B B_{loc}}{k_B T}\right)-\cos \phi \right]\label{energy-uncertainty}.
 \end{equation}
Strictly speaking, the treatment of the mutual interactions would require a many body approach.
In fact, in a many-body picture the mutual interactions can be described by the Hamiltonian \cite{Feyn,Bramwell}:
\begin{equation}
H=-J'\sum_{(ij)}\vec{S}_i^{z_i}\cdot \vec{S}_j^{z_j}+Dr_{nn}^3 \sum_{i>j} \frac{\vec{S}_i^{z_i}\cdot \vec{S}_j^{z_j}}{|\vec{r}_{ij}|^3}-\frac{3(\vec{S_i}^{z_i}\cdot\vec{r}_{ij})(\vec{S_j}^{z_j}\cdot\vec{r}_{ij})}{|\vec{r}_{ij}|^5},
\label{manybody}
\end{equation}
where $\vec{S}$ is the spin vector oriented along the local $z_i$ Ising $<111>$ axis, $i$ and $j$ refer to the two sites of the lattice, $\vec{r}$ is the position vector, $D=(\mu_0\mu^2)/(4\pi r_{nn}^3)$, and $r_{nn}$ is the distance between nearest-neighbor spins. The second term of the Hamiltonian (Eq.\,\ref{manybody}) embodies the magnetic energy associated with the interaction between a single magnetic moment and its nearest neighbors, since the dipolar interaction is short range. Hence, in our analysis of Eq.\,\ref{energy-uncertainty} we consider that the local magnetic field perceived by a certain magnetic moment is associated with the resultant magnetic field generated by the various neighboring magnetic moments in its immediate surrounding, cf.\,Hamiltonian \ref{manybody}.
Upon analyzing Eq.\,\ref{energy-uncertainty}, it is clear that $\Delta E$ is minimized when $\cos \phi \rightarrow$ 1. Such a condition would indicate a maximum alignment of the magnetic moments because of the local magnetic field. Thus, in order to  increase the temperature of the system adiabatically, a quasi-static process is required, i.e., the temporal uncertainty $\Delta t$ should be maximized while $\Delta E$ is minimized. However, still considering Eq.\,\ref{energy-uncertainty}, for $B_{loc}$ = 0 $\Rightarrow \Delta E$ = 0, indicating that in the absence of mutual interactions the uncertainty principle would be violated. In the same way,  the absence of the mutual interactions would imply in the violation of the third law of Thermodynamics, as pointed out previously \cite{ashcroft}.  The only way to reduce the magnetic energy adiabatically is, however, varying the total angular momentum projection. Such a behavior is naturally expected in the adiabatic magnetization, since $\cos \phi$ is increased. At this point, we recall the magnetic energy, namely $U_{mag}=-\frac{m_J}{\sqrt{J(J+1)}} \mu B_{loc}$. In the adiabatic magnetization process, the projection of $\vec {J}$ on the $z$-axis is increased, while its projections on the $x$ and $y$-axes are reduced and, as a consequence, the magnetic energy decreases, being thus the magnetization of the system increased, since $m_J = -J$, $-J+1$, ..., $J-1$, $J$.
This is the principle of work behind the adiabatic magnetization of the mutual interactions here proposed. The magnetic field increment  $\Delta B_{loc}$  of the mutual interaction that emerges in the adiabatic magnetization is associated with the increased projection of $\vec{J}$ along the $z$-axis. In general terms,
in the adiabatic magnetization process (Fig.\,\ref{Fig-8}) of the mutual interactions, the temperature is adiabatically increased from $T_1$ to $T_2$ and, in order to do so, the magnetic energy needs to compensate the temperature variation in order to hold the entropy constant. Thus, we can write:
\begin{equation}
\frac{\mu_B B_{loc}}{k_B T_1} = \frac{\mu_B \sqrt{B_{loc}^2 + {\Delta B_{loc}}^2}}{k_B T_2},
\end{equation}
where $\Delta B_{loc}$ is the magnetic field increment that will emerge into the system to compensate the adiabatic increase of temperature. It is clear that such $\Delta B_{loc}$ refers to the adiabatic magnetization employing only the mutual interactions of the system. Then, $\Delta B_{loc}$ can be easily determined by:
\begin{equation}
\Delta B_{loc} = B_{loc}\sqrt{\frac{T_2^2}{T_1^2} - 1}.
\end{equation}
It is worth emphasizing that such adiabatic magnetization process can only be performed in the temperature range where the magnetic energy associated with the mutual interactions are not overcome by the thermal fluctuations. In the case of real paramagnets ($B_{loc} \simeq$  0.01\,T), as discussed previously, the typical temperature onset of the mutual interactions relevance is $T \leq$ 6\,mK. After magnetizing the system adiabatically, if the temperature is further increased such magnetization will be released when the thermal energy equals the corresponding magnetic energy associated with $\Delta B_{loc}$. Thus, the entropy is recovered to its original configuration, i.e., $\Delta B_{loc} =$ 0\,T, making this process a single shot closed-cycle, which can be restarted. Although we have provided a particular numerical example in Fig.\,\ref{Fig-8}, namely $T_1$ ($T_2$) = 2.0 (2.1)\,mK, such process can be carried out  between any $T_1$ and $T_2$ within the temperature range of relevance of $B_{loc}$. The fascinating aspect behind this process is that no external magnetic field is required in order to perform the adiabatic magnetization, being only the mutual interactions of the system employed.

\vspace{0.3cm}
\textbf{\emph{iv}) On the role of the mutual interactions in other physical systems.}

An analogous situation to the mutual interactions in paramagnets can also be found for the Bose-Einstein condensation (BEC). Strictly speaking, BEC should occur at $T$ = 0\,K, but due to the presence of finite interactions between particles BEC takes place at $T \neq$ 0\,K \cite{boseeinstein}, being the energy of the system equivalent to the chemical potential itself and all particles occupy the same energy level. The fact that interactions between particles prevent an ideal BEC corresponds to a similar physical situation of how the mutual interactions in insulating paramagnets prevent the existence of ideal paramagnets. Also, in the frame of BEC in magnetic insulators \cite{Sachdev,Giamarchi}, where dimers formed by pairs of spin $S$ = 1/2 with a ground state $S$ = 0 (spin-singlet) and $S$ = 1 (triplet) bosonic excitations, the interdimer interaction plays the role analogous to the mutual interaction
in insulating paramagnets. Indeed, the interdimer interaction is the key ingredient for the stabilization of a long-range magnetic ordering \cite{boseeinstein}. Yet, in dipolar spin-ice systems \cite{Bramwell} the dipole-dipole interactions are relevant, so that we consider an ``effective nearest-neighbor energy scale'' composed by two distinct contributions, namely nearest-neighbor exchange energy between $<111>$ Ising magnetic moments and a nearest neighbor dipolar energy scale \cite{Bramwell}. The ``spin-spin'' interactions in such systems can be nicely tuned via doping, which in turn enables to change the distance between magnetic moments \cite{Rafael}. For spins arranged in a pyrochlore structure of corner-sharing tetrahedra spin-ice, we have a residual entropy originating from the degeneracy of the ``disordered'' ground-state.

\section*{Conclusions}
Our findings suggest the absence of zero-field quantum criticality in the presence of mutual interactions. Interactions will eventually, as $T \rightarrow$ 0\,K and $B \rightarrow$ 0\,T, break down the spin gas scheme, so spontaneous magnetization will emerge at some very low-temperature with $B$ = 0\,T.
We have shown that the magnetic Gr\"uneisen parameter for \emph{any} paramagnet with mutual interactions will never diverge as $T \rightarrow $ 0\,K and $B \rightarrow$ 0\,T. As a consequence, zero-field quantum criticality cannot be inferred. As a matter of fact, we are faced to a many body interactions problem, cf.\,Hamiltonian \ref{manybody}.
Our analysis, based on the magnetic Gr\"uneisen parameter and entropy arguments, although validated by the results obtained for the textbook Brillouin-paramagnet and for the proposed zero-field quantum critical paramagnet $\beta$-YbAlB$_4$, is universal and can be applied to \emph{any} system with mutual interactions. Also, we have discussed the impossibility of achieving absolute zero temperature due to the presence of finite mutual interactions in real paramagnets regarding the connection between $\Gamma_{mag}$ and the canonical definition of temperature. Given the recent advances in the investigation of model systems employing cold atoms \cite{Bloch}, our findings suggest that the mutual interactions can be mimicked by tuning the Hubbard on-site Coulomb repulsion $U$, as reported in Ref.\cite{Achim}. Yet, we have proposed the concept of adiabatic magnetization employing solely the mutual interactions. We anticipate that the proof of concept regarding adiabatic magnetization of the mutual interactions can be achieved, with some technical efforts, in any laboratory where temperatures of a few mK are attainable precisely. The role of the mutual interactions in other physical systems like Bose-Einstein condensates, Bose-Einstein condensation of magnons and spin-ice were also discussed.

\section*{Methods}
\vspace{0.3cm}
\textbf{Softwares.}
All the calculations presented in this work were performed employing the software Wolfram Mathematica\textsuperscript{\textregistered} Version 11.  Figures \ref{Fig-1} and \ref{Fig-6} were created from scratch using the software Adobe Illustrator\textsuperscript{\textregistered} Version CC 2017. Figures \ref{Fig-2}, \ref{Fig-4}, \ref{Fig-5},  \ref{Fig-7}, and \ref{Fig-8} were plotted using the software OriginPro\textsuperscript{\textregistered} Version 2018 based on data set generated employing the software Wolfram Mathematica\textsuperscript{\textregistered} Version 11. Figure \ref{Fig-3} was plotted using the 3D plot function of the software Wolfram Mathematica\textsuperscript{\textregistered} Version 11. In Figs.\,\ref{Fig-2}, \ref{Fig-4} (lower panel) and \ref{Fig-5}, $T =$ 5\,mK was chosen arbitrarily due to the numerical impossibility of computing $S$ and $\Gamma_{mag}$ exactly at $T = $ 0\,K. The same holds true for $T =$ 1\,mK in Fig.\,\ref{Fig-5}.

\vspace{0.3cm}
\textbf{Derivation of the observables considering $B_{loc}$ in the Brillouin paramagnet.}
The partition function $Z$ for the Brillouin paramagnet \cite{blundell} reads:
\begin{equation}
Z(B,T) = \sum_{m_J = -J}^{J} \textmd{exp}\left(\frac{\mu_B g m_J B}{k_B T}\right),
\label{Z1}
\end{equation}
converging to the expression:
\begin{equation}
Z(B,T) = \frac{\sinh{\left[\left(1+\frac{1}{2J}\right)\frac{\mu_B J g B}{k_B T}\right]}}{\sinh{\left(\frac{\mu_B J g B}{k_B T}\right)}}.
\end{equation}
Considering $J$ = 1/2 and $g$ = 2 we have:
\begin{equation}
Z(B,T) = \frac{\sinh\left({2\frac{\mu_B B}{k_B T}}\right)}{\sinh{\left(\frac{\mu_B B}{k_B T}\right)}},
\end{equation}
which in turn is simplified as:
\begin{equation}
Z(B,T) = 2\cosh{\left(\frac{\mu_B B}{k_B T}\right)}.
\end{equation}
Considering the existence of a local magnetic field $B_{loc}$ which can  vectorially be added to the external magnetic field $B$, the resultant magnetic field is $B_r = \sqrt{B^2 + {B_{loc}}^2}$, cf.\,previous discussions. Replacing $B$ by $B_r$ in Eq.\,\ref{Z1}, the partition function is now given by:
\begin{equation}
Z(B,T) = 2\cosh{\left(\frac{\mu_B \sqrt{B^2+{B_{loc}}^2}}{k_B T}\right)}.
\label{Z2}
\end{equation}
Thus, the corresponding Helmholtz free energy reads:
\begin{equation}
F(B,T) = -k_B T\ln{\left[2\cosh{\left(\frac{\mu_B \sqrt{B^2+{B_{loc}}^2}}{k_B T}\right)}\right]}.
\label{F2}
\end{equation}
From Eq.\,\ref{F2}, the entropy $S = -\left(\partial F/ \partial T\right)_B$ is:
\begin{equation}
S(B,T) = -\frac{\sqrt{B^2+{B_{loc}}^2} \mu_B}{T}\tanh{\left(\frac{\sqrt{B^2+{B_{loc}}^2} \mu_B}{k_B T}\right)} + k_B\textmd{ln}\left[2\cosh{\left(\frac{\sqrt{B^2+{B_{loc}}^2} \mu_B}{k_B T}\right)}\right].
\label{S1}
\end{equation}
Equation \ref{S1} is exactly the same as Eq.\,\ref{entropyBrillouin} when $B$ is replaced by $B_r$. We thus have shown that only replacing $B$ as $B_r$ in the expressions for the Brillouin paramagnet is consistent, since computing the observables from the partition function considering $B_r$ instead of $B$ provides the same results.
It is worth mentioning that we have performed all the magnetic field derivatives of the entropy with respect to $B$ instead of $B_r$, since $B_{loc}$ was considered constant in our analysis. If we compute the magnetic field derivatives of $S$ with respect to $B_r$ instead of $B$, there would be a difference between such expressions by a factor of $B$/$\sqrt{B^2 + {B_{loc}}^2}$, which is 1 when $B_{loc} = 0$. Interestingly enough, the calculation of $\Gamma_{mag}$ is be affected by this aspect. Considering $\left(\partial S/ \partial B\right)_T$ in Eq.\,\ref{Gamma}, $\Gamma_{mag}$ reads:
\begin{equation}
\Gamma_{mag} = \frac{B}{(B^2 + {B_{loc}}^2)},
\end{equation}
which becomes $\Gamma_{mag}$ = 1/$B$ when $B_{loc} = 0$.
Yet, when $\left(\partial S/ \partial B_r\right)_T$ is considered in Eq.\,\ref{Gamma}, $\Gamma_{mag}$ is given by:
\begin{equation}
\Gamma_{mag} = \frac{1}{\sqrt{B^2 + {B_{loc}}^2}},
\end{equation}
which is also $\Gamma_{mag}$ = 1/$B$ when $B_{loc} = 0$ \cite{prbmce}. The very same argument can be used when dealing with the calculations of $\Gamma_{mag}$ for $\beta$-YbAlB$_4$ \cite{yosuke} taking $B_{loc}$ into account.

\newpage
\section*{Acknowledgements}
M. de S. acknowledges financial support from the S\~ao Paulo Research Foundation -- Fapesp (Grants No. 2011/22050-4 and 2017/07845-7), National Council of Technological and Scientific Development -- CNPq (Grants No.\,302498/2017-6), and T.U.V.S.O.T.E. ACS acknowledges CNPq (Grant No.\,305668/2018-8). This work was partially granted by Coordena\c c\~ao de Aperfei\c coamento de Pessoal de N\'ivel Superior - Brazil (Capes) - Finance Code 001 (Ph.D. fellowship of L.S. and I.F.M.). The Boston University Center for Polymer Studies is supported by NSF
Grants PHY-1505000, CMMI-1125290, and CHE-1213217, and by DTRA Grant
HDTRA1-14-1-0017. A.C.S. acknowledges financial support from the National
Council of Technological and Scientific Development -- CNPq
(Grant No.\,305668/2018-8).

\section*{Author contributions}
L.S., G.O.G., and I.F.M. carried out the calculations and generated the figures. L.S. and M. de S. wrote the paper with contributions from A.C.S., R.E.L., G.O.G., and I.F.M. All authors revised the manuscript. M. de S. conceived and supervised the project.

\section*{Additional Information}
The authors declare no competing interests.

\selectlanguage{english}
\FloatBarrier

\begin{thebibliography}{10}

\bibitem{Mathur} Mathur, N. D. et al. Magnetically mediated superconductivity in heavy fermion compounds. \emph{Nature} \textbf{394,}  39–43 (1998).

\bibitem{Geibel} Gegenwart, P. et al.
Magnetic-Field Induced Quantum Critical Point in
YbRh$_2$Si$_2$.  \emph{Phys. Rev. Lett.} \textbf{89,} 056402 (2002).

\bibitem{kittel} Kittel, C. \emph{Introduction to Solid State Physics 8th ed.} (John Wiley \& Sons, Hoboken, 2005).

\bibitem{ashcroft} Ashcroft, N. W. and Mermim, N. D. \emph{Solid State Physics} (Saunders College Publishing, Orlando, 1976).

\bibitem{weiss} Weiss, P. L'hypoth\`ese du champ mol\'eculaire et la propri\'et\'e ferromagn\'etique. \emph{J. Phys. Theor. Appl.} \textbf{6,} 661-690 (1907).

\bibitem{linuspauling} Pauling, L. A theory of ferromagnetism. \emph{Proc. Nat. Acad. Sci. U. S. A.} \textbf{39,} 551-560 (1953).

\bibitem{cryophysics} Mendelssohn, K. \emph{Cryophysics} (Interscience Publishers, London, 1960).

\bibitem{absolutezero} Mendelssohn, K. \emph{The Quest for Absolute Zero} (World University Library, London, 1966).

\bibitem{madelung} Madelung, O. \emph{Introduction to Solid-State Theory} (Springer-Verlag Berlin Heidelberg, New York, 1978).

\bibitem{vanvleck} van Vleck, J. H. – Nobel Lecture. NobelPrize.org. Nobel Media AB 2019. Fri. 27 Sep 2019. $<$https://www.nobelprize.org/prizes/physics/1977/vleck/lecture/$>$.

\bibitem{ralph}
Baierlein, R. \emph{Thermal Physics} (Cambridge University Press, Cambridge, 1999).

\bibitem{blundell}
Blundell, S. \emph{Magnetism in Condensed Matter} (Oxford University Press, Oxford, 2001).

\bibitem{tari} Tari, A. \emph{The Specific Heat of Matter at Low Temperatures} (Imperial college press, London, 2003).

\bibitem{pathria}
Pathria, R. K. \emph{Statistical Mechanics 2nd ed.} (Butterworth-Heinemann, Oxford, 1996).

\bibitem{feynmann} Feynmann, R., Leighton, R. B. \& Mathew, S. \emph{Lecture on Physics Vol. II and III} (Addison-Wesley, Palo Alto, 1964).

\bibitem{pobel} Pobell, F. \emph{Matter and Methods at Low Temperatures} (Springer-Verlag Berlin Heidelberg, New York, 2007).

\bibitem{fazekas} Fazekas, P. \emph{Lecture Notes on Electron Correlation and Magnetism} (World scientific publishing, Singapore, 1999).

\bibitem{Putzke}
Putzke, C. et al. Anomalous critical fields in quantum critical superconductors. \emph{Nat. Commun.} \textbf{5,} 5679- (2014).

\bibitem{Huang}
Huang, C. L. et al. Anomalous quantum criticality in an itinerant ferromagnet. \emph{Nat. Commun.} \textbf{6,} 8188 (2015).

\bibitem{kanoda}
Isono, T. et al. Quantum criticality in an organic spin-liquid insulator $\kappa$-(BEDT-TTF)$_2$Cu$_2$(CN)$_3$. \emph{Nat. Commun.} \textbf{7,} 13494 (2016).

\bibitem{cuprates}
Michon, B. et al. Thermodynamic signatures of quantum criticality in cuprate superconductors. \emph{Nature} \textbf{567,} 218-222 (2019).

\bibitem{zhu}
Zhu, L., Garst, M., Rosch, A. \& Si, Q. Universally diverging Gr\"uneisen parameter and the magnetocaloric effect close to quantum critical points. \emph{Phys. Rev. Lett.} \textbf{91,} 066404 (2003).

\bibitem{gegenwart}
Gegenwart, P. Classification of materials with divergent magnetic Gr\"uneisen parameter. \emph{Philos. Mag.} \textbf{97,} 3415-3427 (2017).

\bibitem{barto}
Bartosch, L., de Souza, M. \& Lang, M. Scaling theory of the Mott transition and breakdown of the Gr\"uneisen scaling near a finite-temperature critical end point. \emph{Phys. Rev. Lett.} \textbf{104,} 245701
  (2010).

\bibitem{epj}
de Souza, M. et al. Gr\"uneisen parameter for gases and superfluid helium. \emph{Europ. J. of Phys.} \textbf{37,} 055105 (2016).

\bibitem{stanley}
Gomes, G., Stanley, H. E. \& de Souza, M. Enhanced Gr\"uneisen parameter in supercooled water. \emph{Sci. Rep.} \textbf{9,} 12006 (2019).

\bibitem{PNAS} Dasa, D., Gnidaa, D., Wi{\'s}niewskia, P., \&  Kaczorowski, D. Magnetic field-driven quantum criticality in
antiferromagnetic CePtIn$_4$. Proceedings of the National Academy of Sciences, 10.1073/pnas.1910293116 (2019).

\bibitem{mgarst}
Garst, M. \& Rosch, A. Sign change of the Gr\"uneisen parameter and magnetocaloric effect near quantum critical points. \emph{Phys. Rev. B} \textbf{72,} 205129 (2005).

\bibitem{prbmce}
Gomes, G. et al. Magnetic Gr\"uneisen parameter for model systems. \emph{Phys. Rev. B} \textbf{100,} 054446 (2019).

\bibitem{anders} Smith, A. Who discovered the magnetocaloric effect? Warburg, Weiss, and the connection between magnetism and heat. \emph{Eur. Phys. J. H} \textbf{38,} 507–517 (2013).

\bibitem{Moya} Moya, X., Kar-Narayan, S., Marthur, N. D. Caloric materials near ferroic phase transitions. \emph{Nat. Mat.} \textbf{13}, 439-450 (2014).

\bibitem{heavyfermion}
Gegenwart, P., Si, Q. \& Steglich, F. Quantum criticality in heavy-fermion metals. \emph{Nature Phys.} \textbf{4,} 186-197 (2008).

\bibitem{sakai}
Sakai, A. et al. $T$/$B$ scaling without quasiparticle mass divergence: YbCo$_2$Ge$_4$. \emph{Phys. Rev. B} \textbf{94,} 041106 (2016).

\bibitem{yosuke}
Matsumoto, Y. et al. Quantum criticality without tuning in the mixed valence compound $\beta$-YbAlB$_4$. \emph{Science} \textbf{331,} 316-319 (2011).

\bibitem{deguchi}
Deguchi, K. et al. Quantum critical state in a magnetic quasicrystal. \emph{Nature Mater.} \textbf{11,} 1013-1016 (2012).

\bibitem{ramires2012}
Ramires, A. et al. $\beta$-YbAlB$_4$: A critical nodal metal. \emph{Phys. Rev. Lett.} \textbf{109,} 176404 (2012).

\bibitem{tomita2015}
Tomita, T. et al. Strange metal without magnetic criticality. \emph{Science} \textbf{349,} 506-509 (2015).

\bibitem{kuga2008}
Kuga, K. et al. Superconducting properties of the non-Fermi-liquid system $\beta$-YbAlB$_4$. \emph{Phys. Rev. Lett.} \textbf{101,} 137004 (2008).

\bibitem{nakatsuji2008}
Nakatsuji, S. et al. Superconductivity and quantum criticality in the heavy-fermion system $\beta$-YbAlB$_4$. \emph{Nature Phys.} \textbf{4,} 603-607 (2008).

\bibitem{coleman2016}
Coleman, P. Theory perspective: SCES 2016. \emph{Philos. Mag.} \textbf{97,} 3527-3543 (2017).

\bibitem{tomita2015b}
Tomita, T. et al. Unconventional quantum criticality in $\beta$-YbAlB$_4$ detached from its magnetically ordered phase. \emph{Phys. Proc.} \textbf{75,} 482-487 (2015).

\bibitem{Mermin}
Mermin, N. D. \& Wagner, H. Absence of ferromagnetism or antiferromagnetism in one- or two-dimensional isotropic Heisenberg models. \emph{Phys. Rev. Lett.} \textbf{17,} 1113 (1966).

\bibitem{Wolf} Wolf, W. P.  Cooling by adiabatic magnetization.
  \emph{Phys. Rev.} \textbf{115,} 1196 (1959).

\bibitem{odom}
Odom, B. et al. New measurement of the electron magnetic moment using a one-electron quantum cyclotron. \emph{Phys. Rev. Lett.} \textbf{97,} 030801 (2006).

\bibitem{Reif} Reif, F. \emph{Fundamentals of Statistical and Thermal Physics} (Waveland Press, Long Grove, 1965).

\bibitem{alberto}
Guimar\~aes, A. P. \emph{Magnetism and Magnetic Resonance in Solids} (John Wiley \& Sons, New York, 1998).

\bibitem{macaluso}
Macaluso, R. T. et al. Crystal structure and physical properties of polymorphs of LnAlB$_4$ (Ln = Yb, Lu). \emph{Chem. Mater.} \textbf{19,} 1918-1922 (2007).

\bibitem{matsumoto2012}
Matsumoto, Y. et al. $T$/$B$ scaling of magnetization in the mixed valent compound $\beta$-YbAlB$_4$. \emph{J. Phys.: Conf. Ser.} \textbf{391,} 012041 (2012).

\bibitem{kuga2018}
Kuga, K. et al. Quantum valence criticality in a correlated metal. \emph{Science Adv.} \textbf{4,} eaao3547 (2018).

\bibitem{Purcell}
Purcell, E. M. \& Pound, R. V. A nuclear spin system at negative temperature. \emph{Phys. Rev.} \textbf{81,} 279 (1951).

\bibitem{Si} Wu, J., Zhu, L., Si, Q. Entropy accumulation near quantum critical points: effects beyond hyperscaling.  \emph{J. Phys. Conf. Ser.} \textbf{273}, 012019 (2011).

\bibitem{Matsukawa}
Matsukawa, S. et al. Pressure-driven quantum criticality and $T$/$H$ scaling in the icosahedral Au–Al–Yb approximant. \emph{J. Phys. Soc. Japan} \textbf{85,} 063706 (2016).

\bibitem{luquinha} Squillante, L. \& de Souza, M. unpublished results.

\bibitem{Feyn} Feynman R. P., \emph{Statistical Mechanics} (Addison-Wesley, Massachusetts, 1998).

\bibitem{Bramwell} Bramwell, S. T., Gingras, M. J.-P. Spin ice state in frustrated magnetic pyrochlore materials. \textit{Science} \textbf{294}, 1495-1501 (2001).

\bibitem{baxter} Baxter, R. J. \emph{Exactly Solved Models in Statistical Mechanics} (Academic Press, London, 1982).

\bibitem{boseeinstein} G\"{o}rlitz, A. et al. Realization of Bose-Einstein condensates in lower dimensions. \emph{Phys. Rev. Lett.} \textbf{87}, 130402 (2001).

\bibitem{Sachdev} Sachdev, S., Quantum magnetism and criticality. \emph{Nat. Phys.} \textbf{4}, 173-185 (2008).

\bibitem{Giamarchi} Giamarchi T., R\"uegg, C. \& Tchernyshyov, Bose-Einsten condensation in magnetic insulators. \emph{Nat. Phys.} \textbf{4}, 198-204 (2008).

\bibitem{Rafael} Lau, G. C. et al. Zero-point entropy in stuffed spin-ice. \emph{Nat. Phys.} \textbf{2}, 249-253 (2006).

\bibitem{Bloch} Braun, S. et al. Negative absolute temperature for
motional degrees of freedom. \emph{Science} \textbf{339,} 52-55 (2013).

\bibitem{Achim} Rapp, A., Mandt, S. \& Rosch, A. Equilibration rates and negative absolute temperatures for ultracold atoms
in optical lattices.   \emph{Phys. Rev. Lett.} \textbf{105,} 220405 (2010).
\end{thebibliography}
\end{document}